\documentclass[%
 reprint,
 superscriptaddress,
 amsmath,amssymb,
 aps,
]{revtex4-1}

\usepackage{graphicx}%
\usepackage{dcolumn}%
\usepackage{bm}%
\usepackage{float}
\usepackage{graphicx}
\usepackage[utf8]{inputenc}
\usepackage{enumerate}
\usepackage{tabularx}
\usepackage[margin=2.5cm, bottom=0.75in, a4paper]{geometry}
\usepackage{changepage}
\usepackage{listings}	%
\usepackage{amsmath}
\usepackage{amssymb}
\usepackage{enumitem}	%
\usepackage{caption}	%
\usepackage{amsthm}	%
\usepackage{todonotes}	%
\usepackage{marginnote}	%
\usepackage{framed}	%
\usepackage{siunitx}	%
\usepackage{slashed}	%
\usepackage{mathrsfs}	%
\usepackage{makecell}	%
\usepackage{tikz}
\usetikzlibrary{arrows,shapes,trees,decorations.markings, calc,tikzmark}
\usepackage[compat=1.1.0]{tikz-feynman}
\usepackage{mathtools}
\newcounter{tipcount}
\makeatletter
\def\thm@space@setup{%
  \thm@preskip=0.5in plus 0.1in minus 0.2in
  \thm@postskip=\topsep %
}
\makeatother

\newcommand{\pd}[2]{\frac{\partial #1}{\partial #2}} 
\newcommand{\pdd}[2]{\frac{\partial^2 #1}{\partial #2^2}}

\renewcommand{\v}[1]{{\ensuremath{\boldsymbol{\mathbf{#1}}}}} %

\newcommand{\ket}[1]{\left| #1 \right>} %
\newcommand{\bra}[1]{\left< #1 \right|} %

\newcommand{\avg}[1]{\left< #1 \right>} %
\newcommand{\abs}[1]{\left| #1 \right|} %
\DeclareMathOperator{\Tr}{Tr} %
\DeclareMathOperator{\tr}{tr}
\renewcommand{\Re}{\operatorname{Re}}
\renewcommand{\Im}{\operatorname{Im}}
\newcommand{\up}{\uparrow}
\newcommand{\down}{\downarrow}
\newcommand{\pathint}[1]{\int\!\mathcal{D}[#1]\,}	%

\newcommand{\rom}[1]{%
  \textup{\uppercase\expandafter{\romannumeral#1}}  %
}
\newcommand*\chem[1]{\ensuremath{\mathrm{#1}}}

\DeclareMathOperator{\Pf}{Pf}

\setitemize{noitemsep}
\let\OLDthebibliography\thebibliography
\renewcommand\thebibliography[1]{
  \OLDthebibliography{#1}
  \setlength{\parskip}{0pt}
\setlength{\itemsep}{0pt plus 0.3ex}
}

\theoremstyle{plain}

\theoremstyle{definition}

\theoremstyle{remark}

\begin{document}
\title{Derivation of a Ginzburg-Landau free energy density containing mixed gradient terms of a $p+ip$ superconductor with spin-orbit coupling}
\author{Fredrik Nicolai Krohg}
\author{Asle Sudb\o}
\affiliation{\footnotesize Department of Physics, Norwegian University of Science and Technology, NO-7491, Trondheim, Norway}
\affiliation{\footnotesize Center for Quantum Spintronics, Department of Physics, Norwegian University of Science and Technology, NO-7491, Trondheim, Norway}
\date{\today}%

\begin{abstract}
  A Ginzburg-Landau free energy for a superconducting chiral p-wave order parameter is derived from a two-dimensional tight binding lattice model with weak spin-orbit coupling included as a general
  symmetry-breaking field. Superconductivity is
  accounted for by a BCS-type nearest neighbor opposite-spin interaction where we project the potential onto the $p$-wave irreducible representation of the square lattice symmetry group and assume this
  to be the dominating order.
  The resulting free energy contains kinetic terms that mix components of the order parameter as well as directional gradients --- so called mixed gradient terms --- as a virtue of the symmetry of the order
  parameter. Spin-orbit coupling and electron-hole anisotropy lead to additional contributions to the coefficients of these terms, increasing the number of necessary phenomenological
  parameters by one compared to previous work, and leading to an increase in the coefficient measuring Fermi surface anisotropy for Rashba spin-orbit coupling in the continuum limit.
\end{abstract}
\maketitle
\section{Introduction}

Spin-orbit coupling (SOC) couples the spin of the electron to its momentum which splits spin-degenerate electronic bands, and is a recurring theme in many novel superconducting systems.
If \chem{SrTiO_3} is slightly doped with \chem{Ca}, there is a region in the temperature-versus-carrier concentration phase diagram where
superconductivity and ferroelectricity coexist, and where the material has broken spatial inversion symmetry --- a key cause of SOC \cite{Gabay17,Rischau17}.
When SOC is a significant factor,
the associated symmetry of the superconductivity is often of an unconventional character. In this context, ``unconventional'' means superconductivity where the order parameter does not have the usual
spin-singlet $s$-wave pairing symmetry \cite{SigristUeda91}. One example is the one-atom layer of \chem{Tl-Pb} compound on a \chem{Si(111)} surface studied by \cite{Matetskiy15}.
This system exhibits
$2$D superconductivity at a critical temperature $T_c\sim2.25\text{K}$ followed by a Berezinskii-Kosterlitz-Thouless (BKT) transition and has Rashba SOC leading to a maximum splitting of spin bands by
$\sim250\text{meV}$. In this case the superconductivity is argued to be non-conventional because the average distance between Cooper pairs is larger than the
Ginzburg-Landau coherence length.

Another example is the $2$D electron liquid in the celebrated \chem{LaAlO_3/SrTiO_3} interface (for a review see Ref.~\cite{Gabay16}). By using a back-gate to apply an
electric potential across the interface, which tunes carrier density, $T_c$ can be increased to $\sim300\text{mK}$ \cite{Caviglia08}. In a certain region, tuning this gate-voltage affects the Rashba spin-orbit coupling dramatically --- reaching values of
$10\text{meV}$. This region also seems to be correlated to where superconductivity develops \cite{Caviglia10}. The unconventional symmetry resulting from large Rashba SOC is evident 
from the critical field parallel to the interface being much larger than what would be expected from the Pauli-limit \cite{Reyren09}.

Finally, it should be mentioned that it was initially the discovery of superconductivity in the heavy fermion system \chem{CePr_3Si} \cite{Bauer04,Bauer042} that helped intensify research efforts into non-centrosymmetric
superconductors. This system exhibits an increase in critical magnetic field compared to the Pauli limit, as well as suppression of superconductivity by non-magnetic impurities.
Other lines of evidence for the unconventional character of the order parameter include indications of line-nodes in the superconducting gap from penetration depth \cite{Bonalde05},
and thermal conductivity measurements \cite{Izawa05}, among others. For a more thorough overview of non-centrosymmetric systems, see Ref.~\cite{Smidman17}.

In this paper, the Ginzburg-Landau free energy density is derived for a %
$2$D square lattice with spin-orbit coupling where a chiral $p_x+ip_y$ symmetry is assumed to describe the dominating pairing channel.
This particular pairing state has attracted much
attention because of its topological properties, which include the existence of topologically protected Majorana edge states as well as Majorana bound states in the core regions of half integer vortices
\cite{Sarma06}. In the context of superfluidity, $p$-wave pairing is realized as the $A$-phase in \chem{^3He} \cite{Leggett75} and has long been hypothesized to be the dominant superconducting
pairing symmetry in \chem{Sr_2RuO_4} \cite{Luke98, Mackenzie03, Kim17}. 

The vortex structure of a phenomenological Ginzburg-Landau theory for a $2$D chiral $p$-wave pairing symmetry \cite{SigristUeda91, Agterberg98} was studied using numerical simulations in \cite{AsleGaraud16}.
A magnetic field breaks the degeneracy between the two components of the order parameter so that one becomes dominant, while the other
only exists close to topological defects like vortices. 
The simulations found that the superconducting vortices tend to arrange themselves in a square lattice of single-quantized vortices
when the magnetic field is very close to the upper critical field, however for slightly lower field strengths the phase diagram is dominated by a triangular lattice consisting of double-quanta vortices,
which are coreless. The relative angular momentum between the dominant and sub-dominant component of the order parameter determines the kinds of vortices possible in the system and originates in the
structure of the mixed gradient terms in the GL free energy. These terms also drive the sub-dominant component \cite{Heeb99}.
An interesting question is therefore what physical parameters influence the phenomenological coefficients of these types of terms.
Mixed gradient terms have also been found in a multi-component GL theory for a $s+is$ pairing state derived through the Eilenberger equations for quasiclassical propagators \cite{garaud16}.
This state breaks time-reversal symmetry, similarly to the chiral $p$-wave state, and is found to exist in a doped four-band model for iron-pnictides \cite{Boker17}.
In this GL theory however, the terms could be eliminated by a simple spin-rotation in contrast to the $p$-wave case.

Expressions for Ginzburg-Landau theory coefficients for general order parameters have previously been derived assuming either pairing in the normal BCS spin-basis and ignoring spin-orbit coupling,
or by pairing in a single spin-orbit split non-degenerate band \cite{Samokhin04}.
Additionally, GL theory has been derived for a superconductor with $p$-wave symmetry and a coexistent ferromagnetic state \cite{DahlSudbo07}. The derivations in this paper will largely follow
the methods used in these two references.

The difference between the current paper and \cite{Samokhin04} is that spin-orbit coupling is considered a symmetry-breaking field on the ordered state when deriving the GL theory.
The spin-orbit coupling strength is assumed to be small compared to the Debye cutoff frequency. 
A similar system was considered in \cite{Hugdal18} where the spin-orbit coupling strength was assumed to be small relative to a Zeeman field. A pairing
state with $p$-wave symmetry in the diagonalized bands was discovered as a result of a Kohn-Luttinger type interaction coming from the transformation of a repulsive $U$ Hubbard-model to the new bands.
In the present case, the interaction is assumed to give rise to a chiral $p+ip$ pairing symmetry in the non-diagonal spin-bands.
This leads to a number of additional terms in the generalized effective mass compared to the limit of zero spin-orbit coupling.

The paper is organized as follows: In Sec.~\ref{sec:model} the model is introduced, first in terms of the single particle properties in Sec.~\ref{sec:model:SPP}, and then
in Sec.~\ref{sec:model:pairing} the pairing interaction is presented with a brief justification. 
A sketch of how the Ginzburg-Landau free energy was derived is given in Sec.~\ref{sec:free} and its form reduced to
the same as in \cite{AsleGaraud16}. The contributions from spin-orbit coupling to the phenomenological coefficients are finally discussed in Sec.~\ref{sec:summ}.
Details of the calculations are relegated to the appendices.

\section{Tight binding model}
\label{sec:model}
\subsection{Single Particle Problem}
\label{sec:model:SPP}
The system is modeled as a two-dimensional square lattice, which has symmetry group $C_{4v}$, where fermions can exist at each lattice site. In the clean limit there is no disorder in the system implying
that the Fourier transformed single-particle Hamiltonian is diagonal in wave-vectors $\v{k}$. Including antisymmetric spin-orbit coupling \cite{Smidman17} by spin-dependent 
hopping between lattice sites, the single-particle
Hamiltonian can be written
\begin{equation}
  \hat{H}_0 = \!\!\!\sum_{\substack{s_1s_2=\up\down\\\v{k}}}\!\!\!\big[\epsilon(\v{k}) + \v{\gamma}(\v{k})\cdot\v{\sigma}\big]_{s_1s_2}c_{\v{k}s_1}^\dagger c_{\v{k}s_2},
  \label{eq:model:singleParticleHam}
\end{equation}
where $\v{\sigma}$ consists of Pauli-matrices, $c_{\v{k}s}$ is the annihilation operators for a fermion with wave-vector $\v{k}$ and spin $s$, and the sum over $\v{k}$ runs over the first Brillouin zone.
Hermiticity of the Hamiltonian implies that $\epsilon(\v{k})$ and $\v{\gamma}(\v{k})$ are real. Time-reversal symmetry implies the restrictions
$\epsilon(\v{k}) = \epsilon(-\v{k})$ and $\v{\gamma}(\v{k}) = -\v{\gamma}(-\v{k})$. If parity symmetry is enforced $\v{\gamma}(\v{k})$ vanishes and this vector is hence identified with the parity breaking
antisymmetric spin-orbit coupling. The Hamiltonian in Eq.~\eqref{eq:model:singleParticleHam} becomes diagonal by a unitary transformation to the helicity basis given by
\begin{equation}
  \v{a}_{\v{k}} = \frac{1}{\sqrt{2}}
  \begin{pmatrix}
	\frac{i\hat{\gamma}^y-\hat{\gamma}^x}{\sqrt{1-\hat{\gamma}^z}}e^{i\phi_+} & \frac{i\hat{\gamma}^y-\hat{\gamma}^x}{\sqrt{1+\hat{\gamma}^z}}e^{i\phi_-}\\
	-\sqrt{1-\hat{\gamma}^z}e^{i\phi_+} & \sqrt{1+\hat{\gamma}^z}e^{i\phi_-}
  \end{pmatrix}^\dagger\v{c}_\v{k},
  \label{eq:model:helicityTransform}
\end{equation}
where $\phi_\pm$ are arbitrary phases and \\$\hat{\gamma}^i = \gamma^i(\v{k})/\abs{\v{\gamma}(\v{k})}$, assuming $\v{\gamma}$ has some non-zero component in the $xy$-plane in spin space,
or a similar transformation if $\v{\gamma}\|\hat{e}_z$ (c.f. Appendix~\ref{app:diagonalization}). The eigenvalues of the single-particle Hamiltonian in this basis are denoted
\begin{equation}
  \epsilon^h_{\v{k}} = \epsilon(\v{k}) + h\abs{\v{\gamma}(\v{k})}.
  \label{eq:model:eigenvalues}
\end{equation}

\subsection{Pairing interaction}
\label{sec:model:pairing}
To include $p$-wave superconductivity in the model, an attractive BCS-type weak-coupling interaction is introduced between electrons given by
 \begin{equation}
   \begin{split}
	 \hat{V} = -&\frac{1}{2}\sum_{\v{kk}'\v{q}}\sum_{s_1s_2s_1's_2'}V_{\v{kk}',s_1s_2s_1's_2'}\\
	 \times&c_{\frac{\v{q}}{2}+\v{k}s_1}^\dagger c_{\frac{\v{q}}{2}-\v{k}s_2}^\dagger c_{\frac{\v{q}}{2}-\v{k}'s_2'} c_{\frac{\v{q}}{2}+\v{k}'s_1'},
   \end{split}
  \label{eq:model:pwaveIrrepBCSInteraction}
\end{equation}
for
\begin{equation}
  V_{\v{kk}',s_1s_2s_1's_2'} = V_b\sum_{m=1}^{d_b}d_{\v{k},s_1s_2}^{(b_m)}\big(d_{\v{k}',s_1's_2'}^{(b_m)})^\ast,
  \label{eq:model:pwaveIrrepBCSInteraction:coeff}
\end{equation}
where $d_{\v{k},s_1s_2}^{(b_m)}$ are coefficients for basis vectors for the $d_b$ dimensional irreducible representation $b$. These basis vectors are odd and linear in $\v{k}$, i.e. a $p$-wave like momentum
dependence in the continuum limit. Since superconductivity is introduced
in the spin-basis, it is assumed that the spin-orbit coupling is sufficiently weak compared to the superconducting energy scale for this pairing between opposite momentum fermions to be valid, i.e.
spin-orbit coupling is treated as a symmetry-breaking field on the superconducting state \cite{Smidman17}.

The exact forms of the basis vectors are found in the process of proving that such an interaction exists for the square lattice. This is done by finding the possible eigenvectors for a general two-particle Hermitian
operator $\hat{V}$ that has eigenvectors consisting of pairs of particles with opposite momentum.
The eigenspace of a Hermitian operator can be separated into irreducible spaces that are representations of the symmetry group of the lattice. By expanding in the spin-momentum basis
of the two-particle Hilbert space, any such eigenvector $\ket{d}$ can be written
\begin{equation}
  \ket{d} = \sum_{\v{k},s_1s_2}d_{s_1s_2}(\v{k})\ket{\v{k},s_1}\ket{-\v{k},s_2}.
  \label{eq:model:eigenvectorExpansion}
\end{equation}
The eigenvectors will also include a cutoff function $f_c(\epsilon_\v{k})$ since the attractive interaction is assumed to only exist on the Fermi surface. This cutoff function is implicit in the notation for $d_{s_1s_2}(\v{k})$.
If the coefficient $d_{s_1s_2}(\v{k})$ is odd in $\v{k}$, then because of the fermionic particle exchange symmetry and because it is periodic in reciprocal lattice vectors it can be
expanded in terms of lattice vectors $\v{R}$ as
\begin{equation}
  d_{s_1s_2}(\v{k}) = \frac{1}{\sqrt{N}}\sum_{\v{R}}\big(\v{\beta}_{\v{R}}\sin(\v{R}\cdot\v{k})\cdot\v{\sigma}i\sigma^y)_{s_1s_2}.
  \label{eq:model:eigenvectorForm}
\end{equation}
These general vectors are then projected down on the space consisting of basis vectors of a particular irreducible representation (irrep.) $b$ of interest by the projection operators \cite{Inui90,Otnes96}
\begin{equation}
  P^{(b)}_{ll} = \frac{d_b}{|C_{4v}|}\sum_{g\in C_{4v}}D_{ll}^{(b)}(g)^\ast g:
  \label{eq:model:projectionOperators}
\end{equation}
where $D^{(b)}_{ll}$ are matrices of the irrep., $g:$ denotes transformation of a vector by the group element $g$, and the index $l$ runs over the dimension $d_b$ of the irrep. The group $C_{4v}$ contains
one two-dimensional irrep. $E$. Projecting down on this irrep. and assuming the eigenspace of $\hat{V}$ only is constructed from nearest neighbour sites
yields a vector space constructed from the orthonormal basis vectors given by the spin-momentum coefficients
\begin{subequations}
  \begin{align}
	\begin{split}
	  d_{s_1s_2}^{(E_y)}(\v{k}) &= -\frac{\hat{\v{z}}}{\sqrt{N}}\sin k_y \cdot \big(\v{\sigma}i\sigma^y\big)_{s_1s_2} \\
	  &\equiv \v{d}^{(E_y)}(\v{k})\cdot(\v{\sigma}i\sigma^y)_{s_1s_2},
	\end{split}\label{eq:model:basisVec1}\\
	\begin{split}
	  d_{s_1s_2}^{(E_x)}(\v{k}) &= \phantom{-}\frac{\hat{\v{z}}}{\sqrt{N}}\sin k_x \cdot \big(\v{\sigma}i\sigma^y\big)_{s_1s_2}\\
	  &\equiv \v{d}^{(E_x)}(\v{k})\cdot(\v{\sigma}i\sigma^y)_{s_1s_2}.
	\end{split}\label{eq:model:basisVec2}
  \end{align}
  \label{eq:model:basisVectors}
\end{subequations}
These are $p$-wave basis vectors since they are linear in $\v{k}$ in the continuum limit. Note that the assumptions of a single 2D square lattice implies that basis vectors that have $\v{k}$ dependencies
with components in the $\hat{e}_z$-direction are neglected.
When $\hat{V}$ is expanded in its eigenvector basis it is therefore possible that it has a channel consisting of the eigenvectors in Eq.~\eqref{eq:model:basisVectors} and it has been proved that
Eq.~\eqref{eq:model:pwaveIrrepBCSInteraction} is a possible interaction.

This $p$-wave channel interaction could originate as the dominant channel of a simpler interaction. As an example, consider the attractive nearest neighbor interaction
\begin{equation}
  \hat{V} = -\frac{V}{2}\sum_{\avg{i,j}}\sum_{s=\up\down}c_{i,s}^\dagger c_{j,-s}^\dagger c_{j,-s} c_{i,s},
  \label{eq:model:nnInteraction}
\end{equation}
which could be considered an effective one-band model from a reduction of a multiband system \cite{Otnes99}.
Finding basis vectors in the eigenspace of nearest neighbor interactions analogous to the irrep. $E$, $\hat{V}$ becomes diagonal in this basis, and can be written on the form of 
Eq.~\eqref{eq:model:pwaveIrrepBCSInteraction}, but with coefficient
\begin{equation}
  \begin{split}
	V_{\v{kk}',s_1s_2s_1's_2'} = V\Big[&\sum_{a=A_1,B_1}\psi^{(a)}_{s_1s_2}(\v{k})\big(\psi^{(a)}_{s_1's_2'}(\v{k}')\big)^\ast\\
	+ &\sum_{m=x,y}d^{(E_m)}_{s_1s_2}(\v{k})\big(d^{(E_m)}_{s_1's_2'}(\v{k}')\big)^\ast\Big]
  \end{split}
  \label{eq:model:nnInteractionDiagonalized:coeff}
\end{equation}
where $a$ runs over the one-dimensional irreps. $A_1$ and $B_1$ which has basis vectors given by
\begin{align}
  \psi^{(A_1)}_{s_1s_2}(\v{k}) &= \frac{1}{\sqrt{2N}}(\cos k_x + \cos k_y)(i\sigma^y)_{s_1s_2}, \label{eq:model:A1BasisVector}\\
  \psi^{(B_1)}_{s_1s_2}(\v{k}) &= \frac{1}{\sqrt{2N}}(\cos k_x - \cos k_y)(i\sigma^y)_{s_1s_2},
  \label{eq:model:B1BasisVector}
\end{align}
and give the extended $s$-wave and $d$-wave channel respectively.
\section{Derivation of Ginzburg-Landau free energy}
\label{sec:free}
The Ginzburg-Landau coefficients are calculated by deriving the free-energy $F$ of the system described in Section~\ref{sec:model}. This free-energy is defined as $F=-\frac{1}{\beta}\ln Z$, where $Z$ is the partition function
and $\beta$ is inverse temperature. The partition function is defined as $Z = \Tr e^{-\beta(\hat{H}-\mu\hat{N})}$, where $\hat{H} = \hat{H}_0 + \hat{V}$ is the Hamiltonian of the system, $\mu$ is the
chemical potential and $\hat{N}$ is the number operator.
Calculating the trace in the path-integral formalism where the annihilation and creation operators get replaced by Gra\ss mann fields $\xi$ and $\xi^\ast$, the Hubbard-Stratonovich transformation is
preformed on the $p$-wave subspace of the potential $\hat{V}$, while the other subspaces are assumed to be insignificant in the low energy theory. 
Given the potential in Eq.~\eqref{eq:model:nnInteractionDiagonalized:coeff}, this subspace is two-dimensional and its contribution to the partition function can thus be written in terms of a path-integral
over the two complex fields $\eta^{(x)}$ and $\eta^{(y)}$ as
\begin{equation}
  \begin{split}
	e^{S_\text{int}} = \pathint{\eta,\eta^\ast}\exp\Big\{\!-\!\int_0^\beta\!\!\mathrm{d}\tau\sum_{\v{q}m}\Big[\frac{2|\eta_q^{(m)}|^2}{V}& \\
	+ \Big(J_q^{m\;\ast}\eta_q^{(m)} + J_q^m\eta_q^{(m)\;\ast}\Big)\Big]\Big\}&,
  \end{split}
  \label{eq:free:HSTransformation}
\end{equation}
where $J_q^m$ is defined as
\begin{equation}
  J_q^m = \sum_{\v{k}s_1s_2}\big(d_{s_1s_2}^{(E_m)}(\v{k})\big)^\ast \xi_{\frac{\v{q}}{2}-\v{k},s_2}\xi_{\frac{\v{q}}{2}+\v{k},s_1},
  \label{eq:free:HS:J}
\end{equation}
and
\begin{equation}
  S_\text{int} = \frac{V}{2}\int_0^\beta\!\!\mathrm{d}\tau\sum_{\v{q}m}J_q^{m\;\ast}J_q^m.
  \label{eq:free:HS:Sint}
\end{equation}
In these equations both the Gra\ss mann fields $\xi$ and the complex fields $\eta$ are dependent on imaginary-time. The time-dependence in the complex fields $\eta$, which are the order parameters
of the system, is neglected since the goal is a
time-independent Ginzburg-Landau theory, while the time-dependence in the Gra\ss mann-fields are converted to sums over Matsubara frequencies.
The system is assumed to be close to the transition temperature $T_c$ so that the free energy can be expanded to second order in the order parameters after integrating out the fermionic degrees of freedom.
The integration itself is preformed by expressing the part of the exponent with quadratic dependence on fermionic fields as an Hermitian form $\v{\xi}^\dagger\check{G}^{-1}\v{\xi}$ using $4$-component Matsubara vectors
$\v{\xi}$, such that the result depends on the determinant of $\check{G}^{-1}$ by
\begin{equation}
  \begin{split}
	Z_\text{ferm} &= \pathint{\xi,\xi^\ast}e^{-\frac{1}{2}\sum\v{\xi}^\dagger \check{G}^{-1}\v{\xi}}\\
	&= \sqrt{\det\check{G}^{-1}} =  e^{\frac{1}{2}\Tr\ln\check{G}^{-1}}.
  \end{split}
  \label{eq:free:fermionicIntegration}
\end{equation}
The expansion to second order in the order parameter is preformed by splitting $\check{G}^{-1}$ into a diagonal matrix $\check{G}_0^{-1}$ independent of $\eta$ and a matrix $\check{\phi}$ for which each element is proportional
to the order parameter components $\eta^{(a)}$. The logarithm in Eq.~\eqref{eq:free:fermionicIntegration} is then expanded by 
\begin{equation}
  \Tr\ln\check{G}^{-1} = \Tr\ln\check{G}_0^{-1} + \Tr\check{G}_0\check{\phi} - \frac{1}{2}\Tr\check{G}_0\check{\phi}\check{G}_0\check{\phi}.
  \label{eq:free:logarithExpansion}
\end{equation}
The first term is absorbed into the normalization of the path-integral over $\eta$ while the second term vanishes trivially which leaves the contribution of the third term.
The single-particle problem in Eq.~\eqref{eq:model:singleParticleHam} and thus also the spin-orbit coupling is included in this integration over fermionic degrees of freedom.
The order parameter is assumed to be slowly varying in real space, which justifies a gradient expansion. Given these assumptions and approximations, the free energy density in momentum space
takes the form
\begin{equation}
  f_\v{q} = A_{ab}(\eta^{(a)}_\v{q})^\ast\eta^{(b)}_\v{q} + K_{ab,ij}(\eta^{(a)}_\v{q})^\ast\eta^{(b)}_\v{q} q^iq^j,
  \label{eq:free:compressedFreeDensity}
\end{equation}
where the Einstein summation convention has been used to drop the summation over directions $i,j=x,y$ and dimensions of the subspace $a,b=x,y$. 
We call $A_{ab}$ the potential energy tensor while $K_{ab,ij}$ is the generalized effective mass tensor \cite{Samokhin04}. It is worth noting that the same expression is obtained regardless of what choice is made for the phases $\phi_{\pm}$ in Eq.~\eqref{eq:model:helicityTransform},
and whether $\v{\gamma}$ has a component in the $xy$-plane or not.

\subsection{Form of the free energy density tensors}
The potential energy tensor derived in Eq.~\eqref{eq:free:compressedFreeDensity} is given by
\begin{equation}
  A_{ab} = \frac{2\delta_{ab}}{V} - \sum_{\v{k}hh'}d^{ab}\big[1-hh'\big(1-2(\hat{\gamma}^z)^2\big)\big]\chi^{hh'},
  \label{eq:free:Aab}
\end{equation}
where $\chi^{hh'}$ is the Matsubara-frequency sum over Green's functions given by
\begin{equation}
  \chi^{hh'} = \frac{1}{\beta}\sum_n\frac{1}{(i\omega_n-\epsilon^h_\v{k})(-i\omega_n-\epsilon^{h'}_\v{k})},
  \label{eq:free:Rhh}
\end{equation}
and
\begin{equation}
  d^{ab} = \big(\v{d}^{(E_a)}(\v{k})\big)^\ast\cdot\v{d}^{(E_b)}(\v{k}).
  \label{eq:free:dab}
\end{equation}
In Eq.~\eqref{eq:free:Rhh} the chemical potential has been absorbed into the definition of $\epsilon_\v{k}^h$. $h,h'\in\{\pm\}$ and is used to reference the two different helicity-bands when written as exponentials,
while used as $\pm1$ when written as factors.
Since the only $\v{k}$-dependencies in this sum are in the Fermi energies, it is invariant
with respect to symmetry transformations. This means that the momentum-sum vanishes if $a\neq b$ since the summand then becomes odd with respect to each of the components of $\v{k}$ (c.f.
definition of $\v{d}^{(E_a)}(\v{k})$ in Eq.~\eqref{eq:model:basisVectors}).

The generalized effective mass tensor in Eq.~\eqref{eq:free:compressedFreeDensity} can be expressed as
\begin{equation}
  \begin{split}
	K_{ab,ij} = &\frac{1}{8}\sum_{\v{k}hh'}d^{ab}\Big\{\big[hh'\big(1-2(\hat{\gamma}^{z})^2\big)-1\big]\chi^{hh'}_{ij}\\
	&+ 2h'h\chi^{hh'}g_{ij}\Big\},
  \end{split}
  \label{eq:free:Kab}
\end{equation}
where
\begin{equation}
  \begin{split}
	g_{ij} = \;&\partial_i\hat{\v{\gamma}}\cdot\partial_j\hat{\v{\gamma}} - 2\partial_i\hat{\gamma}^z\partial_j\hat{\gamma}^z\\
	- \big(&\hat{\v{\gamma}}\cdot\partial_i\partial_j\hat{\v{\gamma}}-2\hat{\gamma}^z\partial_i\partial_j\hat{\gamma}^z\big),
  \end{split}
  \label{eq:free:gij}
\end{equation}
and
\begin{equation}
  \begin{split}
	\chi^{hh'}_{ij} %
	= -\frac{1}{\beta}\sum_n\bigg\{\pd{}{\epsilon}\frac{1}{i\omega_n-\epsilon^h_\v{k}}\pd{}{\epsilon}\frac{1}{-i\omega_n-\epsilon^{h'}_\v{k}}v_i^hv_j^{h'}&\\
	- \Big(\pdd{}{\epsilon}\frac{1}{i\omega_n-\epsilon^h_\v{h}}\Big)\frac{1}{-i\omega_n-\epsilon^{h'}_\v{k}}v_i^hv_j^h&\\
	\phantom{=}-\Big(\pd{}{\epsilon}\frac{1}{i\omega_n-\epsilon^h_\v{k}}\Big)\frac{1}{-i\omega_n-\epsilon^{h'}_\v{k}}m_{h\,ij}^{-1}\bigg\}&\\
	+ h\leftrightarrow h'.\phantom{\Big(\pd{}{\epsilon}\frac{1}{i\omega_n-\epsilon^h_\v{k}}\Big)}\qquad\qquad\;\;\,&
  \end{split}
  \label{eq:free:RhhijExpression}
 \end{equation}
 The inverse effective mass of the $h$ band is given by 
 \begin{equation}
   m_{h\,ij}^{-1} = \frac{\partial^2\epsilon^h_\v{k}}{\partial k^i\partial k^j} = m_{ij}^{-1} + h\partial_i\partial_j|\v{\gamma}|,
   \label{eq:free:mhij}
 \end{equation}
 while the $h$-band Fermi velocity is given by 
 \begin{equation}
   v_i^h = \pd{}{k^i}\epsilon^h_\v{k} = v_i + h\partial_i|\v{\gamma}|.
   \label{eq:free:vhi}
 \end{equation}

\subsection{Approximation of free-energy density tensors in terms of Fermi surface averages}

More useful expressions can be obtained for $A_{ab}$ and $K_{ab,ij}$ by expressing the sums over momenta $\v{k}$ as averages over energy surfaces defined as
\begin{equation}
  \avg{(\;\cdot\;)}_\xi \equiv \frac{1}{N_0(\xi)}\sum_\v{k}(\;\cdot\;)\delta(\epsilon(\v{k})-\xi),
  \label{eq:free:energyAvg}
\end{equation}
where $N_0(\xi)$ is the density of states at energy $\xi$. Including the chemical potential in the definition of $\epsilon(\v{k})$, the Fermi surface is obtained at $\xi=0$.
Let $h\big(\v{k},\epsilon(\v{k})\big)$ be a generic summand in one of the $\v{k}$-sums, with an explicit $\epsilon(\v{k})$ dependence. The momentum-sum is exchanged for a Fermi surface average by inserting an energy
integral over a delta function such that
\begin{equation}
  \begin{split}
	\sum_\v{k}h\big(\v{k},\epsilon(\v{k})\big) &= \int_{-\epsilon_c}^{\epsilon_c}\!\!\!\!\mathrm{d}\xi\; N_0(\xi)\avg{h\big(\v{k},\xi\big)}_\xi\\
	&\approx \avg{\int_{-\epsilon_c}^{\epsilon_c}\!\!\!\!\mathrm{d}\xi\; N_0(\xi)h\big(\v{k},\xi\big)}_0.
  \end{split}
  \label{eq:free:momSumToFermAvg}
\end{equation}
The integral is cut off at $\epsilon_c$ because of the assumption that the interaction potential only allows pairing to happen within some energy shell around the Fermi surface.
The energy average is assumed to be constant over this energy shell such that only the value at $\xi=0$ is considered.
To simplify the resulting integrals, it is assumed that the critical temperature is small compared to the energy cutoff such that
\begin{equation}
  e_c \equiv \frac{\epsilon_c\beta}{\pi} \gg 1.
  \label{eq:free:dimlessCutoff}
\end{equation}
The spin-orbit coupling is additionally assumed to be small compared to the pairing energy range such that $\epsilon_c \gg |\v{\gamma}| \,\;\forall \v{k}$. With these approximations $A_{ab}$ becomes
  \begin{equation}
	\begin{split}
	  A_{ab} = \delta_{ab}\Big[\frac{2}{V} - 8N_F\ln(2e_ce^C)\avg{d^{ab}}_0&\\
	  - 16N_F\avg{d^{ab}\big(1-2(\hat{\gamma}^z)^2\big)f(\rho_\v{k})}_0\Big]&,
	\end{split}
	\label{eq:free:Aab:FermiAvg}
  \end{equation}
while $K_{ab,ij}$ becomes
\begin{widetext}
\begin{equation}
  \begin{split}
	K_{ab,ij} = &\frac{N_F\beta^27\zeta(3)}{(2\pi)^2}\avg{d^{ab}v_iv_j}_0 + N_F'\frac{\ln(2e_ce^C)}{2}\avg{d^{ab}m_{ij}^{-1}}_0 + N_F\Big\{- 2\frac{\beta^2}{\pi^2}\avg{d^{ab}(\hat{\gamma}^z)^2f_3(\rho_\v{k})v_iv_j}_0\\
	+ &\frac{\beta^27\zeta(3)}{(2\pi)^2}\avg{d^{ab}\big(1+(\hat{\gamma}^z)^2\big)\partial_i|\v{\gamma}|\partial_j|\v{\gamma}|}_0 + \frac{\beta}{\pi}\Big\langle d^{ab}\Big[\frac{\rho_\v{k}}{2e_c^2}\big(1+(\hat{\gamma}^z)^2\big) - (\hat{\gamma}^z)^2f_2(\rho_\v{k})\Big]\partial_i\partial_j|\v{\gamma}|\Big\rangle_0\\
	+ &\avg{d^{ab}f(\rho_\v{k})g_{ij}}_0\Big\} +N_F'\Big\{\frac{\beta}{\pi}\Big\langle d^{ab}\Big[f_2(\rho_\v{k})(\hat{\gamma}^z)^2 + \rho_\v{k}\frac{7\zeta(3)}{4}\big(1+(\hat{\gamma}^z)^2\big)\Big](v_i\partial_j|\v{\gamma}|+\partial_i|\v{\gamma}|v_j)\Big\rangle_0\\
	- &\avg{d^{ab}(\hat{\gamma}^z)^2f(\rho_\v{k})m_{ij}^{-1}}_0\Big\}.
  \end{split}
  \label{eq:free:Kab:FermSurfAverages}
\end{equation}
\end{widetext}
The energy range $[-\epsilon_c,\epsilon_c]$ is assumed to be sufficiently small such that $N_0(\xi)\approx N_F + N_F'\xi$ is a good approximation. $N_F=N_0(0)$ is the value
of the density of states at the Fermi level, while $N_F'=N_0'(0)$ is a measure of the particle-hole asymmetry.
The $f$-functions are all convergent sums that vanish in the limit of no spin-orbit coupling defined as
\begin{align}
  f(\rho) &= \Re\sum_{n=0}^\infty\Big(\frac{1}{2n+1+i\rho} - \frac{1}{2n+1}\Big),
  \label{eq:free:f:1}\\
  f_2(\rho) &= \Im\sum_{n=0}^\infty\frac{1}{(2n+1+i\rho)^2},
  \label{eq:free:f:2}\\
  f_3(\rho) &= \Re\sum_{n=0}^\infty\Big(\frac{1}{(2n+1+i\rho)^3} - \frac{1}{(2n+1)^3}\Big).
  \label{eq:free:f:3}
\end{align}
The dimensionless spin-orbit coupling $\rho_\v{k} = \beta|\v{\gamma}|/\pi$. $\zeta(\cdot)$ is the Riemann-zeta function and $C$ in
$e^C$ is the Euler-Mascheroni constant.

\subsection{The limit of zero spin-orbit coupling}

In the limit of zero spin-orbit coupling, the unit vectors $\hat{\v{\gamma}}$ become indeterminate, however the expressions for the free-energy tensors $A_{ab}$ and $K_{ab,ij}$ still have a well defined
limit since all the unit-vector dependencies vanish. To see this, first consider the limit of $|\v{\gamma}|\to0$ of $\chi^{hh'}$. In this limit, the band-energies $\epsilon^h_{\v{k}}\to\epsilon$ such that,
after preforming the sum over Matsubara-frequencies, Eq.~\eqref{eq:free:Rhh} becomes
\begin{equation}
  \lim_{\abs{\v{\gamma}}\to0}\chi^{hh'} = \frac{\tanh\frac{\beta\epsilon}{2}}{2\epsilon} \equiv S\big[\epsilon(\v{k})\big].
  \label{eq:free:Rhh:zeroSOC}
\end{equation}
Since $\chi^{hh'}$ becomes independent of $h$, $\hat{\gamma}^z$ vanishes under the sum over $h$ and $h'$ in Eq.~\eqref{eq:free:Aab}, and leaves
\begin{equation}
  \begin{split}
	\lim_{\abs{\v{\gamma}}\to0}A_{ab} &= \frac{2\delta_{ab}}{V} - 4\sum_\v{k}d^{ab}S\big[\epsilon(\v{k})\big]\\
	&= \delta_{ab}\Big[\frac{2}{V} - 8I\avg{d^{ab}}_0\Big],
  \end{split}
  \label{eq:free:Aab:zeroSOC}
\end{equation}
for the energy-integral
\begin{equation}
  I = \int_{-\epsilon_c}^{\epsilon_c}\!\!\!\mathrm{d}\xi\;N_0(\xi)S[\xi] \approx N_F\ln(2e_ce^C).
  \label{eq:free:Samokhin:I}
\end{equation}
This corresponds to the $A_{ab}$ calculated in \cite{Samokhin04} if $V\to2V$ and $8d^{ab} = \tr[\Psi^\dagger_a\Psi_b]$.

In $R_{ij}^{hh'}$, the limit reduces $v_i^h\to v_i$ and $m_{h\;ij}^{-1}\to m_{ij}^{-1}$ as well as the previously mentioned limit of Fermi energies $\epsilon^h_\v{k}\to\epsilon$ such that
\begin{equation}
  \begin{split}
	\lim_{\abs{\v{\gamma}}\to0}R_{ij}^{hh'} &= v_iv_j\frac{1}{\epsilon}\pdd{}{\epsilon}\Big(\epsilon S[\epsilon]\Big) + m_{ij}^{-1}\pd{}{\epsilon}S[\epsilon]\\
	&\equiv 4S_2[\epsilon]v_iv_j + 2S_1[\epsilon]m_{ij}^{-1}.
  \end{split}
  \label{eq:free:Rhhij:zeroSOC}
\end{equation}
Since $R_{ij}^{hh'}$ is independent of $h$ and $h'$ in the zero spin-orbit limit, the second line in Eq.~\eqref{eq:free:Kab} as well as the parenthesis in the first line vanish under the $hh'$-sum.
Inserting the above expression for $\lim_{\abs{\v{\gamma}}\to0}R_{ij}^{hh'}$ into $K_{ab,ij}$ and converting to Fermi surface averages yields
\begin{equation}
  \begin{split}
	\lim_{\abs{\v{\gamma}}\to0}K_{ab,ij} &= -\sum_\v{k}d^{ab}\big(2S_2[\epsilon]v_iv_j + S_1[\epsilon]m_{ij}^{-1}\big)\\
	&= -2\avg{v_iv_jd^{ab}}_0I_2 - \avg{m_{ij}^{-1}d^{ab}}_0I_1,
  \end{split}
  \label{eq:free:Kab:zeroSOC}
\end{equation}
for the integrals \cite{Samokhin04}
\begin{align}
  I_1 &= \int_{-\epsilon_c}^{\epsilon_c}\!\!\!\mathrm{d}\xi N_0(\xi)S_1[\xi] \approx -\frac{N_F'}{2}\ln(2e_ce^C),
  \label{eq:free:Samokhin:I1}\\
  I_2 &= \int_{-\epsilon_c}^{\epsilon_c}\!\!\!\mathrm{d}\xi N_0(\xi)S_2[\xi] \approx -N_F\frac{7\beta^2\zeta(3)}{8\pi^2}.
  \label{eq:free:Samokhin:I1}
\end{align}
This corresponds to the result for $K_{ab,ij}$ found in \cite{Samokhin04} if $8d^{ab} = \tr[\Psi^\dagger_a\Psi_b]$.

\subsection{Reduction by symmetries}
\label{sec:free:reduction}
By considering the symmetry of the coefficients $K_{ab,ij}$ and $A_{ab}$, the form of the free energy density $f_\v{q}$ in Eq.~\eqref{eq:free:compressedFreeDensity} can be further restricted. Assuming we have chosen
a proper pseudospin-representation \cite{Smidman17}, the spin-orbit coupling vector $\v{\gamma}(\v{k})$ has the property
\begin{equation}
  \v{\gamma}(\v{k}) = \tilde{R}_g\v{\gamma}(R_g^{-1}\v{k}),
  \label{eq:free:gammaRotation}
\end{equation}
for proper and improper rotations $g$ where $R_g$ is the $3\times3$ rotation matrix, and $\tilde{R}_g = -R_g$ for improper rotations. This relationship leads to the conclusion that 
$(\hat{\gamma}(\v{k})^z)^2$ and $\hat{\v{\gamma}}(\v{k})^2$ are invariant under all $C_{4v}$ symmetries. This implies that
$K_{aa,i\bar{i}}$ and $K_{a\bar{a},ii}$ are both odd with respect to each of the components of $\v{k}$ and thus vanish under the $\v{k}$ sum. Here the notation $\bar{a}$ means
\begin{equation}
  \bar{a} = 
  \begin{cases}
	y & \text{if } a=x\\
	x & \text{if } a=y
  \end{cases}.
  \label{eq:free:barNotation}
\end{equation}
Remember that $a,b,i,j\in\{x,y\}$. Using the symmetries $K_{a\bar{a},ij}=K_{\bar{a}a,ij}$ and $K_{ab,ij}=K_{ab,ji}$, the free energy density can be expressed as \cite{SigristUeda91}
\begin{equation}
  \begin{split}
	f_\v{q} = &-\alpha(|\eta^{(x)}_\v{q}|^2 + |\eta^{(y)}_\v{q}|^2)\\
	&+ \kappa_1\big( |q^x\eta^{(x)}_\v{q}|^2 + |q^y\eta^{(y)}_\v{q}|^2 \big)\\
	&+ \kappa_2\big(|q^y\eta^{(x)}_\v{q}|^2 + |q^x\eta^{(y)}_\v{q}|^2\big) \\
	&+ \kappa_3\big[(q^x\eta^{(x)}_\v{q})(q^y\eta^{(y)}_\v{q})^\ast + \text{h.c.}\big]\\
	&+ \kappa_4\big[(q^y\eta^{(x)}_\v{q})(q^x\eta^{(y)}_\v{q})^\ast + \text{h.c.}\big],
  \end{split}
  \label{eq:free:freeEnergyFinal}
\end{equation}
for coefficients $\alpha=-A_{xx}$, 
$\kappa_1 = K_{xx,xx}$, $\kappa_2 = K_{xx,yy}$ and $\kappa_3=\kappa_4=K_{xy,xy}$. Rotating the coordinate system such that
\begin{equation}
  \begin{pmatrix}
	q^x\\
	q^y
  \end{pmatrix} = 
  \begin{pmatrix}
	\cos\theta & -\sin\theta\\
	\sin\theta & \cos\theta
  \end{pmatrix}
  \begin{pmatrix}
	\tilde{q}^x\\
	\tilde{q}^y
  \end{pmatrix},
  \label{eq:free:rotatedCoordinates}
\end{equation}
defining the chiral basis of the order-parameters as
\begin{equation}
  \begin{pmatrix}
	\eta^+\\
	\eta^-
  \end{pmatrix} = \frac{1}{\sqrt{2}}
  \begin{pmatrix}
	1 & i\\
	1 & -i
  \end{pmatrix}
  \begin{pmatrix}
	\eta^{(E_y)}\\
	\eta^{(E_x)}
  \end{pmatrix},
  \label{eq:free:chiralBasis}
\end{equation}
as well as using dimensionless variables \cite{Agterberg98}, the free-energy density can be further reduced to the form 
\begin{equation}
  \begin{split}
	f_\v{q} = &-(|\eta^{+}_{\tilde{\v{q}}}|^2 + |\eta^{-}_{\tilde{\v{q}}}|^2) + |\tilde{\v{q}}\eta_{\tilde{\v{q}}}^+|^2 + |\tilde{\v{q}}\eta_{\tilde{\v{q}}}^-|^2\\
	&+ \Re\Big\{\Big(e^{i2\theta}(\nu+\Delta)+e^{-i2\theta}(1-\Delta)\Big)\\
	&\phantom{0}\times\Big[\tilde{q}^x\eta^+_{\tilde{\v{q}}}(\tilde{q}^x\eta^-_{\tilde{\v{q}}})^\ast - \tilde{q}^y\eta^+_{\tilde{\v{q}}}(\tilde{q}^y\eta^-_{\tilde{\v{q}}})^\ast\Big]\Big\}\\
	&+\Im\Big\{\Big(e^{-i2\theta}(\nu+\Delta)-e^{i2\theta}(1-\Delta)\Big)\\
	&\phantom{+}\times\Big[\tilde{q}^x\eta^-_{\tilde{\v{q}}}(\tilde{q}^y\eta^+_{\tilde{\v{q}}})^\ast + \tilde{q}^y\eta^-_{\tilde{\v{q}}}(\tilde{q}^x\eta^+_{\tilde{\v{q}}})^\ast\Big]\Big\}.
  \end{split}
  \label{eq:free:dimensionlessFreeEnergyDensity}
\end{equation}
Here the dimensionless parameters $\Delta = 2(\kappa_2 - \kappa_3)/(\kappa_1+\kappa_2)$ and $\nu = (\kappa_1-3\kappa_2)/(\kappa_1+\kappa_2)$.
\cite{Agterberg98}.
In the above expression, the parameter $\Delta$ is new compared to the expression in \cite{Agterberg98}, and is necessary because of the additional contributions
to $K_{ab,ij}$ in Eq.~\eqref{eq:free:Kab:FermSurfAverages} as will be discussed below. Dimensionless variables were introduced by the substitution
\begin{equation}
  \begin{pmatrix}
	\eta\\
	\tilde{\v{q}}
  \end{pmatrix} \to
  \begin{pmatrix}
	\eta/\sqrt{\alpha}\\
	\sqrt{\frac{2\alpha}{\kappa_1+\kappa_2}}\tilde{\v{q}}
  \end{pmatrix}.
  \label{eq:free:dimensionlessVarTransformation}
\end{equation}
Choosing $\theta=0$ and transforming to real-space yields a free-energy density of the form
\begin{widetext}
\begin{equation}
  \begin{split}
	f_\text{GL} = -(|\eta^+|^2 + |\eta^-|^2) + |\v{D}\eta^+|^2 + |\v{D}\eta^-|^2 + \big(\nu+1\big)\Re\Big\{&\Big[D_x\eta^+(D_x\eta^-)^\ast - D_y\eta^+(D_y\eta^-)^\ast\Big]\Big\}\\
	+\big(\nu-1+2\Delta\big)\Im\Big\{&\Big[D_x\eta^-(D_y\eta^+)^\ast + D_y\eta^-(D_x\eta^+)^\ast\Big]\Big\}.\\
  \end{split}
  \label{eq:free:realFreeEnergyDensity}
\end{equation}
\end{widetext}
Here $D_i$ stands for a dimensionless gradient in the $i$-direction in real space and the space-dependence of the order parameter is implicit.
\section{Summary}
\label{sec:summ}

Mixed gradient terms in a Ginzburg-Landau free energy are defined as terms of the form $(D_x\eta^+)^\ast D_y\eta^-$, \cite{Heeb99}, i.e. terms mixing different components and directional gradients.
These terms drive the subdominant component of a chiral $p$-wave superconductor that exists in the core of topological
defects like vortices when a magnetic field breaks the degeneracy between the superconducting components. The core structure of vortices are also influenced by these terms in that the structure of the terms determine
the relative phase of the two order-parameters and thus the different kinds of vortices possible \cite{Heeb99,AsleGaraud16}.
It is evident from the definition that the order parameter needs multiple components for such terms to be present. The number of components of the order parameter depends on the number of dimensions
of the irreducible representations that the pairing interaction furnishes.
If the symmetry group contains a two-dimensional irreducible representation and the interaction contains this irrep. as a subspace of its eigenvalue space, then the order parameter associated with
this subspace has two components.
In the weak-coupling BCS framework this discussion is based on, on-site Hubbard interaction on a square $2$D lattice in the clean limit only consists of the one-dimensional
$s$-wave representation. 
If spin-orbit coupling is included as a symmetry-breaking field, then the gap function is rotated in the new basis so that it gains a momentum dependence determined by the SOC spin texture 
\cite{Sergi04,Smidman17}. In the case of Rashba spin-orbit coupling, the transformation is such that the intra-component elements of the gap-function and thus the pairing-amplitude in the
spin-orbit split bands, gains a $p$-wave like momentum dependence. Such systems could thus be called effective $p$-wave superconductors \cite{Loder17}, %
however their topological properties are different
from those of true triplet $p$-wave superconductors and, importantly, the order parameter does not gain additional components.
For the $2$D square lattice, this means that a pairing interaction that acts at least as far as nearest neighbor lattice sites, is necessary for a multi-component order parameter to be present.
This type of interaction was also found to be sufficient to contain a two dimensional subspace given by the $p$-wave irreducible representation basis vectors.

The two mixed gradient terms found in the Ginzburg-Landau free energy $f_\text{GL}$ are determined by two phenomenological parameters $\Delta = 2(\kappa_2-\kappa_3)/(\kappa_1+\kappa_2)$ and 
$\nu = (\kappa_1-3\kappa_2)/(\kappa_1+\kappa_2)$ where $\kappa_1 = K_{xx,xx}$, $\kappa_2 = K_{xx,yy}$ and $\kappa_3 = K_{xy,xy}$ for the generalized effective mass tensor $K_{ab,ij}$.

If both SOC and the particle-hole asymmetry is set to zero, and we assume nearest neighbor hopping, 
$K_{ab,ij}$ reduces to 
\begin{equation}
  K_{ab,ij} = \zeta_{ab}\frac{N_F\beta^27\zeta(3)}{(4\pi)^2Nt^2}\avg{v_av_bv_iv_j}_0,
  \label{eq:summ:Kab:reduced}
\end{equation}
where $\zeta_{ab} = (-1)^{\delta_{ab}-1}$.
With this reduction, $\nu$ can be written as
\begin{equation}
  \nu=\frac{\avg{v_x^4}_0-3\avg{v_x^2v_y^2}_0}{\avg{v_x^4}_0+\avg{v_x^2v_y^2}_0},
  \label{eq:summ:nu}
\end{equation}
as in \cite{Agterberg98,AsleGaraud16} and is thus a measure of the Fermi surface anisotropy. 
The coefficient in front of the last mixed gradient term
becomes $\geq0$ and proportional to $\avg{v_x^2v_y^2}_0$. 
It will therefore exist as long as there is superconducting order and the Fermi velocity does not vanish. The coefficient in front of the first mixed gradient term is on the other hand $(\nu+1)$. From
Eq.~\eqref{eq:summ:nu} we see that for a completely anisotropic square Fermi surface, $\nu=-1$ such that the mixed gradient term vanishes. The remaining term can in this case be rotated away by a rotation
of the order parameter components as in \cite{garaud16}.

With the simplification of $K_{ab,ij}$ in Eq.~\eqref{eq:summ:Kab:reduced}, the parameter $\Delta$ becomes $\Delta = -(\nu-1)$ and the form of $f_\text{GL}$ reduces to that of \cite{Agterberg98} except for a minus sign. This discrepancy originates with the choice made for the
basis of the $p$-wave subspace. To get equality, you would simply choose both eigenvectors positive in Eq.~\eqref{eq:model:basisVectors}, which would yield an irreducible representation equivalent to $E$.
Then $K_{ab,ij}$ would reduce in the same way except missing the factor $\zeta_{ab}$ such that $\Delta=0$ and $f_\text{GL}$ would reduce to the same form.

If the particle-hole asymmetry given by $N_F'$ is present, $K_{ab,ij}$ gains a contribution from the Fermi surface average $\avg{v_av_bm_{ij}^{-1}}_0$. For nearest neighbor hopping, $m_{ij}^{-1}$ is diagonal such that
$\kappa_3$ is not affected by it, however because of its contribution to $\kappa_1$ and $\kappa_2$ the terms get re-scaled. In the continuum limit
this leads to increasing coefficients for the mixed gradient terms compared to the normal kinetic terms in the free energy.

In the continuum limit $\nu$ is expected to vanish by Eq.~\eqref{eq:summ:nu}, since the Fermi surface becomes isotropic. However, including Rashba spin-orbit coupling with SOC-vector of the form 
$\v{\gamma} = \alpha(\v{k}_y\hat{e}_x - \v{k}_x\hat{e}_y)$, leads to 
\begin{equation}
  \nu \approx \frac{1}{2}\left( \frac{\alpha}{k_Ft} \right)^2,
  \label{eq:summ:nu:Rashba}
\end{equation}
where $t>0$ is the nearest neighbor hopping amplitude and $k_F$ is the Fermi wave vector magnitude because of the contribution to the $\kappa$ coefficients from the term 
$N_F\avg{d^{ab}g_{ij}}_0$ in $K_{ab,ij}$. From this result we conclude that $\nu$ is no longer only a measure of Fermi surface anisotropy,
but also is a measure of spin-orbit coupling strength --- or alternatively that spin-orbit coupling gives an effective Fermi surface anisotropy. 
The coefficient in front of the last mixed gradient term in Eq.~\eqref{eq:free:realFreeEnergyDensity} now becomes
$1/(1-\nu)$, while the other mixed gradient term coefficient is $1+\nu$ for the choice $\theta=0$.
This shows that in the continuum limit, the mixed gradient terms
become more prominent compared to the normal gradient terms, as the Rashba spin-orbit coupling strength increases.
\begin{acknowledgments}
F. N. K. was supported by an NTNU university grant, and thanks S. Rex for useful discussions. A.S. was supported by the Research Council of Norway through Grant Number 250985, "Fundamentals of Low-dissipative Topological Matter", and Center of Excellence Grant Number 262633, Center for Quantum Spintronics.
\end{acknowledgments}
\appendix

\section{Symmetries of the single particle problem}
Combining the two different spin-options for the annihilation operators in Eq.~\eqref{eq:model:singleParticleHam} in a vector $\hat{\v{c}}_\v{k}$, then under time-reversal $\hat{\theta}$, the operators
transform as \cite{Schober16}
\begin{subequations}
  \begin{align}
	\hat{\theta}\hat{\v{c}}_\v{k}\hat{\theta}^{-1} &= i\sigma^y\hat{\v{c}}_{-\v{k}},\label{eq:app:SPPSymm:timeRevTransform:1}\\
	\hat{\theta}\hat{\v{c}}_\v{k}^\dagger\hat{\theta}^{-1} &= \hat{\v{c}}^\dagger_{-\v{k}}(-i\sigma^y).
	\label{eq:app:SPPSymm:timeRevTransform:2}
  \end{align}
  \label{eq:app:SPPSymm:timeRevTransform}
\end{subequations}
Since $\hat{\theta}$ contains a conjugation operator, the time-reversal of the single-particle Hamiltonian in Eq.~\eqref{eq:model:singleParticleHam} becomes
\begin{equation}
  \begin{split}
	\hat{\theta}\hat{H}_0\hat{\theta}^{-1} &= \sum_{\v{k}}\hat{\theta}\hat{\v{c}}_\v{k}^\dagger\hat{\theta}^{-1}\,(\epsilon(\v{k})^\ast + \v{\gamma}(\v{k})^\ast\cdot\v{\sigma}^\ast)\hat{\theta}\hat{\v{c}}_\v{k}\hat{\theta}^{-1}\\
	&= \sum_\v{k}\hat{\v{c}}^\dagger_{-\v{k}}\big(\epsilon(\v{k})^\ast + \v{\gamma}(\v{k})^\ast\cdot(-i\sigma^y)\v{\sigma}^\ast(i\sigma^y)\big)\hat{\v{c}}_{-\v{k}}\\
	&= \sum_\v{k}\hat{\v{c}}^\dagger_\v{k}\big(\epsilon(-\v{k})^\ast - \v{\gamma}(-\v{k})^\ast\cdot\v{\sigma}\big)\hat{\v{c}}_\v{k}.
  \end{split}
  \label{eq:app:SPPSymm:timeReversedHamiltonian}
\end{equation}
If the Hamiltonian should be time-reversal invariant, then the coefficients must have the symmetries $\epsilon(\v{k}) = \epsilon(-\v{k})^\ast$ and $\v{\gamma}(\v{k}) = -\v{\gamma}(-\v{k})^\ast$. Since
the Pauli-matrices are self-adjoint, taking
the adjoint of $\hat{H}_0$ yields
\begin{equation}
  \hat{H}_0^\dagger = \sum_\v{k}\hat{\v{c}}^\dagger_\v{k}\big(\epsilon(\v{k})^\ast + \v{\gamma}(\v{k})^\ast\cdot\v{\sigma}\big)\hat{\v{c}}_\v{k}.
  \label{eq:app:SPPSymm:adjointHamiltonian}
\end{equation}
If the Hamiltonian should be Hermitian, then the coefficients must satisfy $\epsilon(\v{k}) = \epsilon(\v{k})^\ast$ and $\v{\gamma}(\v{k}) = \v{\gamma}(\v{k})^\ast$.
Time-reversal invariance together with Hermiticity thus implies that the coefficients are real, that $\epsilon(\v{k})$ is even in $\v{k}$ and that $\v{\gamma}(\v{k})$ is odd in $\v{k}$, which were the
symmetries mentioned in Section~\ref{sec:model:SPP}.

\section{Diagonalization of the single particle problem}
\label{app:diagonalization}
It is easily verified through substitution that the basis defined in Eq.~\eqref{eq:model:helicityTransform} diagonalizes the Hamiltonian in Eq.~\eqref{eq:model:singleParticleHam} as long as
$|\hat{\gamma}^z|\neq1$, regardless of whether $\hat{H}_0$ is Hermitian or time-reversal invariant. This means that the same diagonalization is used when $\v{\gamma}$ represents spin-orbit coupling
(time-reversal invariant but not parity invariant), and when it represents an external magnetic field (parity invariant but not time-reversal invariant). The matrix determining the basis in
Eq.~\eqref{eq:model:helicityTransform} is found by solving the characteristic equation of the corresponding linear-algebra problem and finding the normal eigenvectors that correspond to each eigenvalue.

In the case that $\v{\gamma}(\v{k})\|\,\hat{e}_z$ the basis transformation instead reads
\begin{equation}
  \v{a}_\v{k} = \frac{1}{2}
  \begin{pmatrix}
	(1+\hat{\gamma}^z)e^{i\phi_+} & (1-\hat{\gamma}^z)e^{i\phi_-}\\
	(1-\hat{\gamma}^z)e^{i\phi_+} & (1+\hat{\gamma}^z)e^{i\phi_-}
  \end{pmatrix}^\dagger
  \v{c}_\v{k}.
  \label{eq:app:helicityTransformAlternative}
\end{equation}
This results in the same expression for the eigenvalues $\epsilon^h_\v{k} = \epsilon(\v{k}) + h|\v{\gamma}(\v{k})|$ as the basis transformation in Eq.~\eqref{eq:model:singleParticleHam}

\section{Basis vector for the irreducible representation $E$ of $C_{4v}$}

The group of symmetry transformations of the two-dimensional square lattice is denoted $C_{4v}$ in the Sch\"onflies notation or $4mm$ in the abbreviated Hermann-Maugin notation \cite{Inui90}. In
\cite{Inui90} the character table of $C_{4v}$ is as shown in Table~\ref{tab:app:basis:C4vChar}. For the one-dimensional irreps. the matrix elements of the representation is the characters themselves.
For the two-dimensional irrep. $E$, the matrix elements of the representation are given by
\begin{equation}
  \begin{split}
&\begin{aligned}
	D^{(E)}(e) &= 
	\begin{pmatrix}
	  1 & 0\\
	  0 & 1
	\end{pmatrix}\\
	D^{(E)}(C_4^2) &= 
	\begin{pmatrix}
	  -1 & 0\\
	  0 & -1
	\end{pmatrix}\\
	D^{(E)}(C_4) &= 
	\begin{pmatrix}
	  0 & -1\\
	  1 & 0
	\end{pmatrix}\\
	D^{(E)}(C_4^{-1}) &= 
	\begin{pmatrix}
	  0 & 1\\
	  -1 & 0
	\end{pmatrix}
\end{aligned}
\begin{aligned}
  D^{(E)}(\sigma_x) &=
	\begin{pmatrix}
	  -1 & 0\\
	  0 & 1
	\end{pmatrix}\\
	D^{(E)}(\sigma_y) &=
	\begin{pmatrix}
	  1 & 0\\
	  0 & -1
	\end{pmatrix}\\
	D^{(E)}(\sigma_{d_1}) &=
	\begin{pmatrix}
	  0 & -1\\
	  -1 & 0
	\end{pmatrix} \\
	D^{(E)}(\sigma_{d_2}) &=
	\begin{pmatrix}
	  0 & 1\\
	  1 & 0
	\end{pmatrix}
\end{aligned}\\
	  \end{split}
  \label{eq:app:basis:G5Rep}
\end{equation}
This can be verified by calculating the traces of the matrices $\chi^{(E)}(g)$ and showing that they satisfy the condition
\begin{equation}
  \sum_{g\in C_{4v}}|\chi^{(E)}(g)|^2 = |C_{4v}|,
  \label{eq:app:basis:irrepCondition}
\end{equation}
which imply that this is an irreducible representation,
as well as showing that the matrices satisfy the group multiplication-relations for group-elements in $C_{4v}$.

\begin{table}[h]
  \centering
  \begin{tabular}{c !{\vrule width0.8pt} c | c | c | c | c |}
	$C_{4v}$ & $e$ & $C_4^2$ & $2C_4$ & $2\sigma_v$ & $2\sigma_d$ \\ \Xhline{0.8pt}
	$A_1$ & $1$ & $1$ & $1$ & $1$ & $1$ \\ \hline 
	$A_2$ & $1$ & $1$ & $1$ & $-1$ & $-1$ \\ \hline 
	$B_1$ & $1$ & $1$ & $-1$ & $1$ & $-1$ \\ \hline 
	$B_2$ & $1$ & $1$ & $-1$ & $-1$ & $1$ \\ \hline 
	$E$ & $2$ & $-2$ & $0$ & $0$ & $0$ \\ \hline 
  \end{tabular}
  \caption{Character table for the group $C_{4v}$. The first row gives the conjugation classes, while the first column denotes the different irreducible representations. Note that $E$ is the only two-dimensional irreducible representation.}
  \label{tab:app:basis:C4vChar}
\end{table}

Since the goal is to find a basis for this representation $E$ consisting of eigenvectors of the Hermitian operator $\hat{V}$, these basis vectors can be written on the
form of Eq.~\eqref{eq:model:eigenvectorExpansion}, repeated here for convenience:
\begin{equation}
  \ket{d} = \sum_{\v{k},s_1s_2}d_{s_1s_2}(\v{k})\ket{\v{k},s_1}\ket{-\v{k},s_2}.
  \label{eq:app:basis:eigenvectorExpansion}
\end{equation}
This eigenvector-space is projected down on the irreducible subspace of the irreducible representation by the projection operator in Eq.~\eqref{eq:model:projectionOperators}. This operator includes
the symbol $g:$ which means that the state should be transformed by the group-element $g$. For spin-momentum eigenstates, the transformation law is given by \cite{Schober16}
\begin{equation}
  g:\ket{\v{k}',s'} = \sum_s\ket{g\v{k}',s}D_{g\,ss'}
  \label{eq:app:basis:spinMomTransformation}
\end{equation}
for the matrix
\begin{equation}
  D_{g\,ss'} = \delta_{ss'}\cos(\phi/2)-i\hat{\v{u}}\cdot\v{\sigma}_{ss'}\sin(\phi/2)
  \label{eq:app:basis:spinMatrix}
\end{equation}
where the rotation given by the angle and normal vector $(\phi,\,\hat{\v{u}})$ is given by the proper rotation associated with $g$. Transformation of vectors in the product space of two spin-momentum
eigenstates are thus given by
\begin{equation}
  \begin{split}
	g:&\ket{\v{k}'_1,s_1'}\ket{\v{k}'_2,s_2'}\\
	= \sum_{s_1s_2}&\ket{g\v{k}'_1,s_1}\ket{g\v{k}_2',s_2}D_{g\,s_1s_1'}D_{g\,s_2s_2'}.
  \end{split}
  \label{eq:app:basis:spinMomProductTransformation}
\end{equation}
Writing this as an active transformation where the transformation acts on the coefficients of the eigenvectors in Eq.~\eqref{eq:app:basis:eigenvectorExpansion}, results in
\begin{equation}
  g:d_{s_1s_2}(\v{k}) = \sum_{s_1's_2'}D_{g\,s_2s_2'}D_{g\,s_1s_1'}d_{s_1's_2'}(g^{-1}\v{k}).
  \label{eq:app:basis:eigenvectorCoeffTransformation}
\end{equation}
Coefficients that are odd in $\v{k}$ can be written as
\begin{equation}
  d_{s_1s_2}(\v{k}) = \v{d}(\v{k})\cdot(\v{\sigma} i\sigma^y)_{s_1s_2},
  \label{eq:app:basis:oddEigenvectorCoefficients}
\end{equation}
because of fermionic particle exchange asymmetry. The transformation rule in Eq.~\eqref{eq:app:basis:eigenvectorCoeffTransformation} is then simplified to
\begin{equation}
  g:\v{d}(\v{k}) = R(\hat{\v{u}},\phi)\v{d}(g^{-1}\v{k}),
  \label{eq:app:basis:oddEigenvectorCoeffTransformation}
\end{equation}
where $R$ is the conventional $3\times3$ rotation matrix, which shows that $\v{d}$ transforms as a vector. Since the $\v{k}$-dependency of $\v{d}(\v{k})$ must be such that it is invariant with respect to
translations by reciprocal lattice vectors it can be expanded as a Fourier series in the fundamental lattice vectors $\v{R}$ such that
\begin{equation}
  \v{d}(\v{k}) = \frac{1}{\sqrt{N}}\sum_\v{R}\v{\beta}_\v{R}\sin\v{R}\cdot\v{k}.
  \label{eq:app:basis:latticeExpansionOfd}
\end{equation}
Applying the projection operators in Eq.~\eqref{eq:model:projectionOperators} onto $\v{d}(\v{k})$ using the transformation law in Eq.~\eqref{eq:app:basis:oddEigenvectorCoeffTransformation} and the
matrix-elements of the representation given in Eq.~\eqref{eq:app:basis:G5Rep}, the $\hat{x}$ and $\hat{y}$ components of $\v{\beta}$ vanish, leaving
\begin{equation}
  \begin{split}
	P_{ll}^{(E)}\v{d}(\v{k}) = \frac{\hat{\v{z}}}{2\sqrt{N}}\sum_\v{R}\beta^z_\v{R}\big[\sin(\v{R}\cdot\v{k})&\\
	+ (-1)^l\sin(\hat{z}\cdot\v{R}\times\v{k})&\big].
  \end{split}
  \label{eq:app:basis:dProjection}
\end{equation}
This expression implies immediately that the simplest potential that contains a non-vanishing representation $E$ is a nearest neighbor potential where $\v{R}\in\{(0,\pm1), (\pm1,0)\}$.
Inserting these possible lattice vectors $\v{R}$ in the sum $\sum_\v{R}$ in Eq.~\eqref{eq:app:basis:dProjection}, vectors in the projected space can be written
\begin{equation}
  \begin{split}
	P_{ll}^{(E)}\v{d}(\v{k}) = &\frac{\hat{\v{z}}}{2\sqrt{N}}\big[(\beta^z_{(1,0)}-\beta^z_{(-1,0)})\\
	  &\times(\sin  k_x + (-1)^l\sin  k_y ) \\
	+ &(\beta^z_{(0,1)} - \beta^z_{(0,-1)})\\
  &\times(\sin k_y - (-1)^l\sin  k_x)\big].
  \end{split}
  \label{eq:app:basis:dProjectionNearestNeighbor}
\end{equation}
Such vectors can clearly all be written using the basis vectors made up of
\begin{equation}
  \v{d}_\pm(\v{k}) = \hat{\v{z}}\,(\sin k_x \pm \sin k_y).
	\label{eq:app:basis:1stBasis}
\end{equation}
Although this is a basis for the irreducible vector space associated with the irreducible representation $E$, it doesn't transform as the matrices given in Eq.~\eqref{eq:app:basis:G5Rep}. Recall that a
basis $\{b_i\}$ for a representation $D$ transforms according to
\begin{equation}
  g:b_i = \sum_jb_jD_{ji}(g).
  \label{eq:app:basis:groupTransformationOfBasisVectors}
\end{equation}
Instead $\{\v{d}_\pm\}$ transforms like an equivalent representation to the matrices in Eq.~\eqref{eq:app:basis:G5Rep}. This is simply solved by rotating the basis into new basis vectors
\begin{subequations}
  \begin{align}
	\v{d}^{(E_y)}(\v{k}) &= -\hat{\v{z}}\sin k_y,\label{eq:app:basis:unNormalizedBasis:2}\\
	\v{d}^{(E_x)}(\v{k}) &= +\hat{\v{z}}\sin k_x,
	\label{eq:app:basis:unNormalizedBasis:2}
  \end{align}
  \label{eq:app:basis:unNormalizedBasis}
\end{subequations}
which when properly normalized gives the basis set in Eq.~\eqref{eq:model:basisVectors}.
\section{Spectral Decomposition of nearest neighbor interaction}

To find the representations the potential in Eq.~\eqref{eq:model:nnInteraction} consists of, first it is Fourier transformed into
\begin{equation}
  \begin{split}
	\hat{V} = -&\sum_{\v{q}\v{k}\v{k}'s}\tilde{V}(\v{k}-\v{k}')\\
	\times&c_{\frac{\v{q}}{2}+\v{k},s}^\dagger c_{\frac{\v{q}}{2}-\v{k},-s}^\dagger c_{\frac{\v{q}}{2}-\v{k}',-s} c_{\frac{\v{q}}{2}+\v{k}',s},
  \end{split}
  \label{eq:app:spec:FTPotential}
\end{equation}
for
\begin{equation}
  \tilde{V}(\v{k}-\v{k}') = \frac{V}{2N}\sum_\v{\delta}e^{\v{\delta}\cdot(\v{k}-\v{k}')},
  \label{eq:app:spec:FTFormFactor}
\end{equation}
where $\v{\delta}$ sums over nearest neighbor lattice vectors. 
The spectral decomposition of $\hat{V}$ is found by expressing $\hat{V}$ in terms of its eigenvectors. Since $\hat{V}$ is a two-body operator, it is completely determined by the matrix elements
$\bra{\alpha\beta}\hat{V}\ket{\alpha'\beta'}$ where $\ket{\alpha\beta}$ are states in the two-particle Hilbert space. For BCS type potentials this two-particle Hilbert space consists of states where the
particles have opposite momentum and any eigenvector can thus be expanded as in Eq.~\eqref{eq:model:eigenvectorExpansion}. This means that in terms of spin-momentum eigenstates, the potential can be written as
\begin{equation}
  \begin{split}
	\hat{V} = \frac{1}{2}&\sum_{\v{qkk}'}\sum_{s_1s_2s_3s_4}V_{\v{k},\v{k}';\,s_1s_2s_3s_4}\\
	\times&c_{\frac{\v{q}}{2}+\v{k},s_1}^\dagger c_{\frac{\v{q}}{2}-\v{k},s_2}^\dagger c_{\frac{\v{q}}{2}-\v{k}',s_4} c_{\frac{\v{q}}{2}+\v{k}',s_3},
  \end{split}
  \label{eq:app:spec:potentialInMomSpinEigenstates}
\end{equation}
for the matrix elements
\begin{equation}
  \begin{split}
	V_{\v{k},\v{k}';\,s_1s_2s_3s_4} &= \bra{\v{k},s_1}\bra{-\v{k},s_2}\hat{V}\ket{\v{k}',s_3}\ket{-\v{k}',s_4}\\
	&= -2\tilde{V}(\v{k}-\v{k}')\delta_{s_1s_3}\delta_{s_2s_4}\sigma^x_{s_1s_2}.
  \end{split}
  \label{eq:app:spec:momSpinPotentialMatrixElements}
\end{equation}

The space associated with a single eigenvalue can in general be written as a sum
of irreducible spaces where each irreducible space consists of basis vectors forming a basis for an irreducible representation of the symmetry group \cite{Inui90}. If the space consists of several
irreducible representations, these are said to have accidental symmetry since the fact that vectors belonging to two different irreducible spaces have the same eigenvalue is not necessary by symmetry
and thus, in a sense, accidental. Writing the basis vectors for an irreducible representation $\Gamma$ as $\ket{\Gamma,m_\Gamma}$ where $m_\Gamma$ enumerates the dimensions of the irrep., this implies that 
$\{\ket{\Gamma,m_\Gamma}\}$ is a complete orthonormal basis-set. Inserting this complete set on either side of the potential operator, in the space of two-particle states the potential can be represented by
\begin{equation}
  \hat{V} = \sum_\Gamma V_\Gamma\sum_{m=1}^{d_\Gamma}\ket{\Gamma,m_\Gamma}\bra{\Gamma,m_\Gamma},
  \label{eq:app:spec:potentialInIrrepBasis}
\end{equation}
for the eigenvectors
\begin{equation}
  V_\Gamma = \bra{\Gamma,m_\Gamma}\hat{V}\ket{\Gamma,m_\Gamma}.
  \label{eq:app:spec:irrepEigenvalues}
\end{equation}
Note that it does not matter which of the $d_\Gamma$ different basis-vectors one inserts for $m_\Gamma$ since all 
will give the same eigenvalue as long as they are basis vectors in the same irreducible space.
These eigenvalues can then be evaluated by inserting a complete set of spin-momentum eigenstates as
\begin{equation}
  \begin{split}
	V_\Gamma %
	= \sum_{\v{kk}'}\sum_{s_1s_2s_3s_4}&\; V_{\v{k},\v{k}';\,s_1s_2s_3s_4}\;\\
	\times&(d^{(\Gamma,m_\Gamma)}_{\v{k},\,s_1s_2})^\ast d^{(\Gamma,m_\Gamma)}_{\v{k}',\,s_3s_4}.
  \end{split}
  \label{eq:app:spec:irrepEigenvaluesBySpinMom}
\end{equation}
Inserting the irreducible representation basis vectors in Eq.~\eqref{eq:model:B1BasisVector}, \eqref{eq:model:A1BasisVector} and \eqref{eq:model:basisVectors} yields the eigenvalues
\begin{equation}
  V_{A_1} = V_{B_1} = V_{E_x} = V_{E_y} = -V.
  \label{eq:app:spec:eigenvaluesForDifferentIrreps}
\end{equation}
Conversely, Eq.~\eqref{eq:app:spec:potentialInIrrepBasis} may be inserted into Eq.~\eqref{eq:app:spec:momSpinPotentialMatrixElements} such that the spin-momentum eigenstate
matrix elements can be written as
\begin{equation}
  \begin{split}
	V_{\v{k},\v{k}';\,s_1s_2s_3s_4} %
	&= \sum_\Gamma V_\Gamma\sum_{m_\Gamma=1}^{d_\Gamma}d^{(\Gamma, m_\Gamma)}_{\v{k},\,s_1s_2}(d^{(\Gamma,m_\Gamma)}_{\v{k}',\,s_3s_4})^\ast.
  \end{split}
  \label{eq:app:spec:momSpinPotentialIrrepMatrixElements}
\end{equation}
If all the eigenvectors given by irreducible representations have been accounted for, this must reproduce Eq.~\eqref{eq:app:spec:FTPotential}. Inserting the singlet irreducible
irreducible representations with even functions $\psi^{(a)}(\v{k})$ as well as the triplet irreducible basis vectors with odd vector functions $\v{d}^{(E_i)}(\v{k})$ from 
Eq.~\eqref{eq:model:B1BasisVector}, \eqref{eq:model:A1BasisVector} and \eqref{eq:model:basisVectors}, yields
\begin{widetext}
\begin{equation}
  \begin{split}
	\hat{V} &= -V\sum_{\v{qkk}'s}\Big(\sum_{a=A_1,B_1}\psi^{(a)}(\v{k})\psi^{(a)}(\v{k}')^\ast + \sum_{i=x,y} d_z^{(E_i)}(\v{k}) d_z^{(E_i)}(\v{k}')^\ast\Big)\; c_{\frac{\v{q}}{2}+\v{k},s}^\dagger c_{\frac{\v{q}}{2}-\v{k},-s}^\dagger c_{\frac{\v{q}}{2}-\v{k}',-s} c_{\frac{\v{q}}{2}+\v{k}',s}\\
	&= -\frac{V}{N}\sum_{\v{qkk}'s}\tilde{V}(\v{k}-\v{k}')c_{\frac{\v{q}}{2}+\v{k},s}^\dagger\; c_{\frac{\v{q}}{2}-\v{k},-s}^\dagger c_{\frac{\v{q}}{2}-\v{k}',-s} c_{\frac{\v{q}}{2}+\v{k}',s},
  \end{split}
  \label{eq:app:spec:potentialFullCircle}
\end{equation}
\end{widetext}
which indeed is the initial potential presented in Eq.~\eqref{eq:app:spec:FTPotential}.
This shows that Eq.~\eqref{eq:model:nnInteractionDiagonalized:coeff} is the diagonalized form of Eq.~\eqref{eq:model:nnInteraction} and the nearest neighbor interaction thus consists of the irreducible
representations $A_1$, $B_1$ and $E$ which corresponds to extended $s$-wave, $d$-wave and $p$-wave channel respectively.

\section{Integration over Fermions}

The single particle problem Hamiltonian $\hat{H}_0$ defined in Eq.~\eqref{eq:model:singleParticleHam} and interaction potential $\hat{V}$ defined in 
Eq.~\eqref{eq:model:pwaveIrrepBCSInteraction} for $b$ equal to the two-dimensional irreducible representation $E$ of $C_{4v}$ with eigenvectors given in Eq.~\eqref{eq:model:basisVectors}
defines the relevant system. The
finite temperature partition function for this system can then be written as a path-integral over Gra\ss mann fields $\xi$ and $\xi^\ast$ as 
\begin{equation}
  Z = \pathint{\xi^\ast\xi}e^{-S},
  \label{eq:app:int:generalPartitionFunction}
\end{equation}
for the action
\begin{equation}
  \begin{split}
	S = \int_0^\beta\!\!\!\!\mathrm{d}\tau\Big\{&\sum_{\v{k}ss'}\xi_{\v{k},s}^\ast\big(\delta_{ss'}(\partial_\tau + \epsilon(\v{k})) + \v{\gamma}\cdot\v{\sigma}_{ss'}\big)\xi_{\v{k},s'}\\
	- \frac{V}{2}&\sum_{\v{q}m}J_\v{q}^{m\;\ast}J_\v{q}^m\Big\},
  \end{split}
  \label{eq:app:int:action}
\end{equation}
where $J_\v{q}^m$ are defined in Eq.~\eqref{eq:free:HS:J}. By Hubbard-Stratonovich transforming the interaction potential exponential at the expense of introducing new auxiliary complex fields
$\eta_\v{q}^{(m)}$ and $\eta_\v{q}^{(m)\;\ast}$ as in Eq.~\eqref{eq:free:HSTransformation}, the partition function can be factorized into a path-integral over the auxiliary fields and a path-integral over the quadratic fermionic Gra\ss mann fields
by
\begin{equation}
  Z = \pathint{\eta^\ast\eta}e^{-\int_0^\beta\!\!\!\mathrm{d}\tau\sum_{\v{q}m}\frac{2|\eta_\v{q}^{(m)}|^2}{V}}Z_F,
  \label{eq:app:int:afterHubbStratPartitionFunction}
\end{equation}
such that
\begin{equation}
  Z_F = \pathint{\xi^\ast\xi}e^{-S_F}.
  \label{eq:app:int:fermionicPartitionFunction}
\end{equation}
Because of the Hubbard-Stratonovich transformation, the fermionic action $S_F$ now consists of only quadratic combination of Gra\ss mann fields, where one part of it comes from the single particle problem
on the first line of Eq.~\eqref{eq:app:int:action} and the other is proportional with the new complex fields $\eta$. To simplify the calculation, the Gra\ss mann fields are transformed
through Eq.~\eqref{eq:model:helicityTransform} to the helicity-basis in which the single particle Hamiltonian is diagonal. Denoting the unitary matrix in the transformation in Eq.~\eqref{eq:model:helicityTransform},
$U(\v{k})_{sh}$ such that
\begin{equation}
  \xi_{\v{k}s} = \sum_hU(\v{k})_{sh}\zeta_{\v{k}h},
  \label{eq:app:int:helicityBasisGrassmannFields}
\end{equation}
the fermionic action $S_F$ can be written
\begin{equation}
  \begin{split}
	S_F %
	= &\int_0^\beta\!\!\!\mathrm{d}\tau\bigg\{\sum_{\v{k}h}\zeta_{\v{k}h}^\ast(\partial_\tau + \epsilon_\v{k}^h)\zeta_{\v{k}h}\\
	&+ \sum_{\substack{\v{k}_1\v{k}_2\\h_1h_2m}}\Big[\eta^{(m)}_{\v{k}_1+\v{k}_2}\tilde{d}^{(E_m)}_{\v{k}_1\v{k}_2;h_1h_2}\zeta_{\v{k}_1h_1}^\ast\zeta_{\v{k}_2h_2}^\ast\\
	&+ \eta^{(m)\;\ast}_{\v{k}_1+\v{k}_2}\big(\tilde{d}^{(E_m)}_{\v{k}_1\v{k}_2;h_1h_2}\big)^\ast\zeta_{\v{k}_2h_2}\zeta_{\v{k}_1h_1}\Big]\bigg\},
  \end{split}
  \label{eq:app:int:fermionicAction}
\end{equation}
where in the last equality we have inserted the helicity basis and defined the helicity transformed irrep. basis vectors
\begin{equation}
  \begin{split}
	\tilde{d}^{(E_m)}_{\v{k}_1\v{k}_2;h_1h_2} = \sum_{s_1s_2}&d^{(E_m)}_{s_1s_2}\big({\scriptstyle\frac{\v{k}_1-\v{k}_2}{2}}\big)\\
	\times&U(\v{k}_2)_{s_2h_2}^\ast U(\v{k}_1)_{s_1h_1}^\ast.
  \end{split}
  \label{eq:app:int:helicityIrrep}
\end{equation}
The imaginary-time dependence of the $\zeta$ fields is expanded in a series of Matsubara frequencies through the unitary transformation
\begin{equation}
  \zeta_{\v{k}h}(\tau) = \frac{1}{\sqrt{\beta}}\sum_ne^{-i\tau\omega_n}\zeta_{\v{k}hn},
  \label{eq:app:int:matsubaraFermionicFields}
\end{equation}
for $\omega_n = (2n+1)\pi/\beta$. This expansion results in a remaining time-dependence in the auxiliary complex fields $\eta(\tau)$ which is itself transformed into a bosonic Matsubara-frequency 
dependence through the identification
\begin{equation}
  \frac{1}{\beta}\int_0^\beta\!\!\!\mathrm{d}\tau\;\eta^{(m)}_{\v{k}_1+\v{k}_2}(\tau)e^{i\tau(\omega_{n_1} + \omega_{n_2})} = \eta^{(m)}_{\v{k}_1+\v{k}_2,n_1+n_2+1}.
  \label{eq:app:int:matsubaraTransformedComplexFields}
\end{equation}
In the single particle Hamiltonian, this transformation exchanges the $\partial_\tau$ for $-i\omega_{n_2}$. The fermionic action is now written as a bi-linear form
\begin{equation}
  S_F = \frac{1}{2}\sum_{\substack{\v{k}_1\v{k}_2\\n_1n_2}}\v{\zeta}_{\v{k}_1n_1}^\mathrm{T}\check{A}_{\v{k}_1\v{k}_2,n_1n_2}\v{\zeta}_{\v{k}_2n_2},
  \label{eq:app:int:fermAction:bilinearForm}
\end{equation}
through the $4\times4$ matrix $\check{A}$ by collecting the fermionic fields in four-component vectors
\begin{equation}
  \v{\zeta}_{\v{k}n}^\mathrm{T} = (\zeta_{\v{k}+n},\;\zeta_{\v{k}-n},\;\zeta_{\v{k}+n}^\ast,\;\zeta_{\v{k}-n}^\ast).
  \label{eq:app:int:MajoranaVector}
\end{equation}
Since each vector contains all the different Gra\ss mann fields (both the fields $\zeta$ and $\zeta^\ast$), the integral becomes the Pfaffian of the anti-symmetric component of $\check{A}$
\cite{Serre02}. Re-using the notation $\check{A}$ for this anti-symmetric component, the fact that the Pfaffian of an antisymmetric matrix can be expressed as the square root of the determinant of this
matrix \cite{Wimmer12}, is used to write
\begin{equation}
  Z_F = \Pf(\check{A}) = \pm\sqrt{\det(\check{A})} = e^{\frac{1}{2}\Tr\ln\check{A}}.
  \label{eq:app:int:Pfaffian}
\end{equation}
The limit of zero spin-orbit coupling is used to argue that $+$ should be used in front of the square root. 
The fact that exchanging two rows of a matrix leaves the determinant invariant, is then used to write $S_F$ as the familiar sesquilinear form
\begin{equation}
  S_F = \frac{1}{2}\sum_{\substack{\v{k}_1\v{k}_2\\n_1n_2}}\v{\zeta}_{\v{k}_1n_1}^\dagger (\check{G}^{-1})_{\v{k}_1\v{k}_2,n_1n_2}\v{\zeta}_{\v{k}_2n_2},
  \label{eq:app:int:fermAction:sesquilinearForm}
\end{equation}
where the inverse Gor'kov Green's function $\check{G}^{-1}$ is expressed as
\begin{equation}
  \check{G}^{-1} = \check{G}_0^{-1} + \check{\phi}.
  \label{eq:app:int:separationOfMeanFieldAndOrderParameterMatrix}
\end{equation}
The two terms represent the inverse mean field Green's function
\begin{equation}
  \begin{split}
  &(\check{G}_0^{-1})_{\v{k}_1\v{k}_2,n_1n_2} = \delta_{\v{k}_1\v{k}_2}\delta_{n_1n_2}\\
  &\times\begin{pmatrix}
	-i\omega_{n_1} + \epsilon_{\v{k}_1}^+ & 0\\
	0 & -i\omega_{n_1} + \epsilon^-_{\v{k}_1}
  \end{pmatrix}\otimes\sigma^z,
  \label{eq:app:int:meanFieldMatrix}
\end{split}
\end{equation}
and the order-parameter dependent $4\times4$ matrix
\begin{equation}
  \begin{split}
	&(\check{\phi})_{\v{k}_1\v{k}_2,n_1n_2} = 2\sum_m\sum_n\delta_{n,n_1+n_2+1}\\
	&\times
  \begin{pmatrix}
	0 & \eta^{(m)}_{\v{k}_1+\v{k}_2,n}D^{(m)}_{\v{k}_1\v{k}_2}\\
	\eta^{(m)\;\ast}_{\v{k}_1+\v{k}_2, n}D^{(m)\;\dagger}_{\v{k}_2\v{k}_1} & 0
  \end{pmatrix},
  \label{eq:app:int:orderParamMatrix}
\end{split}
\end{equation}
where the $2\times2$ matrix $D^{(m)}_{\v{k}_1\v{k}_2}$ consists of the transformed irrep. basis vectors
\begin{equation}
  (D^{(m)}_{\v{k}_1\v{k}_2})_{h_1h_2} = \tilde{d}^{(E_m)}_{\v{k}_1\v{k}_2;h_1h_2}.
  \label{eq:app:int:transformedBasisVectorMatrix}
\end{equation}
The result in Eq.~\eqref{eq:app:int:Pfaffian} is then expanded to second order in the order parameter through Eq.~\eqref{eq:free:logarithExpansion}. The first term is independent of $\eta$ and is thus
absorbed in the normalization-constant of the $\eta$ path-integral. The second term vanishes when taking the trace, leaving the third term such that
\begin{equation}
  Z_F = e^{-\frac{1}{4}\Tr\check{G}_0\check{\phi}\check{G}_0\check{\phi}}.
  \label{eq:app:int:fermPartiFunction:traceExpandedForm}
\end{equation}
Since $\check{G}_0^{-1}$ is a completely diagonal matrix, its inverse is trivial to find. By simple matrix multiplication and summing over the momentum and Matsubara-frequency indices for the trace, it is
found that
\begin{equation}
  \begin{split}
  \Tr\check{G}_0\check{\phi}\check{G}_0\check{\phi} = 8\sum_{\substack{mm'\;\v{k}\v{k}'\\hh'\;n_1n_2}}\eta^{(m)}_{\v{k}+\v{k}',n_1}\eta^{(m')\;\ast}_{\v{k}+\v{k}',n_1}&\\
  \times\frac{\tilde{d}^{(E_m)}_{\v{k}\v{k}';hh'}\tilde{d}^{(E_{m'})\;\ast}_{\v{k}\v{k}';hh'}}{(i\omega_{n_2}-i\nu_{n_1}+\epsilon^h_\v{k})(i\omega_{n_2}-\epsilon^{h'}_{\v{k}'})}&
  \label{eq:app:int:secondOrderTrace}
\end{split}
\end{equation}
Since the goal is a time-independent Ginzburg-Landau theory, the order parameter is assumed to be time-independent such that $\eta^{(m)}_{\v{k},n} = \delta_{n0}\eta^{(m)}_\v{k}$. Inserting this assumption
back into Eq.~\eqref{eq:app:int:secondOrderTrace} which is inserted into $Z_F$ in Eq.~\eqref{eq:app:int:fermPartiFunction:traceExpandedForm} and then inserting this back into the expression for $Z$ in
Eq.~\eqref{eq:app:int:afterHubbStratPartitionFunction} yields the expression
\begin{equation}
  \begin{split}
  Z = &\pathint{\eta^\ast\eta}\exp\bigg\{-\sum_{\v{q}m}\beta\frac{2|\eta_\v{q}^{(E_m)}|^2}{V} \\
  - &2\sum_{\substack{mm'\,\v{k}\v{k}'\\hh'\,n}}\eta_{\v{k}+\v{k}'}^{(E_m)}\eta_{\v{k}+\v{k}'}^{(E_{m'})\;\ast}\\
  &\times\frac{\tilde{d}^{(E_m)}_{\v{k}\v{k}';hh'}\tilde{d}^{(E_{m'})\;\ast}_{\v{k}\v{k}';hh'}}{(i\omega_{n}+\epsilon^h_\v{k})(i\omega_{n}-\epsilon^{h'}_{\v{k}'})}\bigg\}.
  \label{eq:app:int:secondOrderExpandedPartitionFunction}
\end{split}
\end{equation}
After shifting the momentum indices in the second term by
\begin{equation}
  \begin{split}
	\v{k} &\to \v{q}/2 + \v{k},\\
	\v{k}' &\to \v{q}/2 - \v{k},
  \end{split}
  \label{eq:app:int:momentumShift}
\end{equation}
inserting the expression for $\tilde{d}^{(E_m)}_{\v{k}_1\v{k}_2;hh'}$ from Eq.~\eqref{eq:app:int:helicityIrrep} as well as the elements of the transformation matrices $U(\v{k})_{sh}$ from
Eq.~\eqref{eq:model:helicityTransform}, $Z$ can be re-written in terms of the gap-function \cite{Samokhin04}
\begin{equation}
  \Delta^{s_1s_2}(\v{k},\v{q}) = \sum_m\eta^{(m)}_\v{q}d^{(E_m)}_{s_1s_2}(\v{k}),
  \label{eq:app:int:gapFunction}
\end{equation}
the spin-orbit dependent matrix
\begin{equation}
  u(\v{k})^h_{ss'} = (\sigma^0 + h\hat{\v{\gamma}}\cdot\v{\sigma})_{ss'}
  \label{eq:app:int:effectiveTransformationMatrix}
\end{equation}
and Green's functions
\begin{equation}
  G^h(\v{k},i\omega_n) = (i\omega_n-\epsilon^h_\v{k})^{-1}
  \label{eq:app:int:bandGreensFunctions}
\end{equation}
as
  \begin{equation}
	\begin{split}
	  Z = &\pathint{\eta^\ast\eta}\exp\bigg\{-\sum_{\v{q}m}\beta\frac{2|\eta_\v{q}^{(E_m)}|^2}{V} \\
	  +&\frac{1}{2}\sum_{\substack{\v{k}\v{q}\;s_1s_2\\s_1's_2'}}\Delta^{s_1s_2}(\v{k},\v{q})\Delta^{s_1's_2'}(\v{k},\v{q})^\ast\\
	  &\times\sum_{nhh'}G^h({\scriptstyle \frac{\v{q}}{2}+\v{k}},-i\omega_n)u({\scriptstyle \frac{\v{q}}{2}+\v{k}})^h_{s_1's_1}\\
	  &\qquad\times G^{h'}({\scriptstyle\frac{\v{q}}{2}-\v{k}},i\omega_n)u({\scriptstyle\frac{\v{q}}{2}-\v{k}})^{h'}_{s_2's_2}\bigg\}.
	\end{split}
	\label{eq:app:int:secondOrderExpandedPartitionFunction:GapFunctionVersion}
  \end{equation}
For further development, the center of mass momentum $\v{q}$ of the Cooper pairs is assumed to be small compared to the fundamental lattice constant so that the momentum dependencies in 
Eq.~\eqref{eq:app:int:secondOrderExpandedPartitionFunction:GapFunctionVersion} can be expanded to second order by
\begin{equation}
  \begin{split}
  u({\scriptstyle\frac{\v{q}}{2}\pm\v{k}})^h &\approx \sigma^0 + h\v{\sigma}\cdot\big(\pm \hat{\v{\gamma}}\\
  &+ \frac{\v{q}^i}{2}\partial_i\hat{\v{\gamma}} \pm \frac{\v{q}^i\v{q}^j}{8}\partial_i\partial_j\hat{\v{\gamma}}\big),
  \label{eq:app:int:momExpansion:u}
\end{split}
\end{equation}
and
\begin{equation}
  \begin{split}
  &\sum_nG^h({\scriptstyle\frac{\v{q}}{2}+\v{k}},-i\omega_n)G^{h'}({\scriptstyle\frac{\v{q}}{2}-\v{k}},i\omega_n)\\
  &= \beta\big(\chi^{h'h} + \frac{\v{q}^i}{2}\chi^{h'h}_i + \frac{\v{q}^i\v{q}^j}{8}\chi^{h'h}_{ij}\big),
  \label{eq:app:int:momExpansion:GreensFunctions}
\end{split}
\end{equation}
where the Einstein-summation convention notation has been used for repeated indices and $\partial_i=\partial/\partial\v{k}^i$. $\chi^{hh'}$ and $\chi^{hh'}_{ij}$ are defined as in Eq.~\eqref{eq:free:Rhh} and
\eqref{eq:free:RhhijExpression}, while
\begin{equation}
  \begin{split}
  \chi^{h'h}_i = \lim_{\v{q}\to0}\pd{}{\v{q}^i}\frac{1}{\beta}\sum_n&G^h(\v{q}+\v{k},-i\omega_n)\\
  \times&G^{h'}(\v{q}-\v{k},i\omega_n).
  \label{eq:app:int:Rhhi}
\end{split}
\end{equation}
Inserting these expansions, the resulting expression for $Z$ becomes
\begin{widetext}
  \begin{equation}
	\begin{split}
	  Z = \pathint{\eta^\ast\eta}\exp\bigg\{-&\beta\sum_{\v{q}m}\frac{2|\eta^{(m)}_\v{q}|^2}{V} \\
	  - \frac{\beta}{2}\sum_{\substack{\v{kq}\,mm'\\hh'}}\eta^{(m)}_\v{q}\eta^{(m')\;\ast}_\v{q}\bigg[&\tr\Big[d_\v{k}^{(E_{m'})\;\dagger}\Big(hh'\hat{\v{\gamma}}\cdot\v{\sigma}d_\v{k}^{(E_m)}\hat{\v{\gamma}}\cdot\v{\sigma}^\mathrm{T} - d_\v{k}^{(E_m)}\Big)\Big]\chi^{h'h}\\
	  - &\frac{\v{q}^i}{2}\tr\big[d_\v{k}^{(E_{m'})\;\dagger}\v{\sigma}d_\v{k}^{(E_m)}\big]\cdot\Big(\hat{\v{\gamma}}(h-h')\chi^{h'h}_i + \partial_i\hat{\v{\gamma}}(h+h')\chi^{h'h}\Big)\\
	  + &\frac{\v{q}^i\v{q}^j}{8}\bigg(\tr\Big[d_\v{k}^{(E_{m'})\;\dagger}\Big(hh'\hat{\v{\gamma}}\cdot\v{\sigma}d_\v{k}^{(E_m)}\hat{\v{\gamma}}\cdot\v{\sigma}^\mathrm{T}-d_\v{k}^{(E_m)}\Big)\Big]\chi^{h'h}_{ij}\\
	  + &2hh'\chi^{h'h}\tr\Big[d_\v{k}^{(E_{m'})\;\dagger}\Big(\hat{\v{\gamma}}\cdot\v{\sigma}d_\v{k}^{(E_m)}\partial_i\partial_j\hat{\v{\gamma}}\cdot\v{\sigma}^\mathrm{T} - \partial_i\hat{\v{\gamma}}\cdot\v{\sigma}d_\v{k}^{(E_m)}\partial_j\hat{\v{\gamma}}\cdot\v{\sigma}^\mathrm{T}\Big)\Big]\bigg)\bigg]\bigg\},
	\end{split}
	\label{eq:app:int:partitionFunction:generalMomentumExpansion}
  \end{equation}
\end{widetext}
where $\tr[\cdot]$ is a trace over the spin-indices, and $d^{(E_m)}_\v{k}$ is the matrix in spin-space whose matrix elements are given by $d^{(E_m)}_{s_1s_2}(\v{k})$. The specific form of
$d^{(E_m)}_{s_1s_2}(\v{k})$ given in Eq.~\eqref{eq:model:basisVectors}, leads to considerable simplifications of Eq.~\eqref{eq:app:int:partitionFunction:generalMomentumExpansion} since the corresponding
spin-vectors $\v{d}^{(E_m)}(\v{k})$ are parallel and only retains the $\hat{\v{z}}$-component. Inserting this fact, the partition function reduces to
\begin{widetext}
\begin{equation}
  \begin{split}
	Z = \pathint{\eta^\ast\eta}\exp\bigg\{-\beta\sum_{\v{q}m}\frac{2|\eta^{(m)}_\v{q}|^2}{V} - \frac{\beta}{2}\sum_{\substack{\v{kq}mm'\\hh'}}\eta^{(m)}_\v{q}\eta^{(m')\;\ast}_\v{q}\tr\Big[d_\v{k}^{(E_m)}d_\v{k}^{(E_{m'})\;\dagger}\Big]\bigg[\big(hh'\big[1-2(\hat{\gamma}^z)^2\big]-1\big)\chi^{hh'}&\\
	+ \frac{q^iq^j}{8}\bigg(\Big[hh'\big(1-2(\hat{\gamma}^z)^2\big)-1\Big]\chi^{hh'}_{ij} - 2h'h\chi^{hh'}g_{ij}\bigg)\bigg]\bigg\}.&
  \end{split}
  \label{eq:app:int:partitionFunction:specialMomentumExpansion}
\end{equation}
\end{widetext}

Since the spin trace in Eq.~\eqref{eq:app:int:partitionFunction:specialMomentumExpansion} can be written
\begin{equation}
  \tr\Big[d_\v{k}^{(E_{m'})\;\dagger}d_\v{k}^{(E_m)}\Big] = 2\big(\v{d}^{(E_{m'})}\big)^\ast\cdot\v{d}^{(E_m)},
  \label{eq:app:int:irrepBasisVectorTrace}
\end{equation}
the free-energy tensors $A_{ab}$ and $K_{ab,ij}$ can now be identified from Eq.~\eqref{eq:app:int:partitionFunction:specialMomentumExpansion} since their relation to the partition function is given by
\begin{equation}
  \begin{split}
	Z = \pathint{\eta^\ast\eta}&\exp\Big\{-\beta\sum_\v{q}\Big[A_{ab}(\eta^{(a)}_\v{q})^\ast\eta_\v{q}^{(b)} \\
	+ &K_{ab,ij}(\eta_\v{q}^{(a)})^\ast\eta_\v{q}^{(b)}\v{q}^i\v{q}^j\Big]\Big\}.
  \end{split}
  \label{eq:app:int:partitionFunction:tensorIdentification}
\end{equation}

\section{Energy integrals in Fermi surface averages}

The details of how to obtain the explicit expression for $K_{ab,ij}$ and $A_{ab}$ in Eq.~\eqref{eq:free:Aab:FermiAvg} and \eqref{eq:free:Kab:FermSurfAverages} 
from Eq.~\eqref{eq:free:Aab} and \eqref{eq:free:Kab}, were in large part left out. In this section,
one of the integrals is worked out in detail and the others needed to obtain these expressions will be listed.

To see clearly what part of the generalized mass tensor $K_{ab,ij}$ is dependent on spin-orbit coupling and which is not, the summation over $h'$ in Eq.~\eqref{eq:free:Kab}
is preformed to yield the expression
\begin{equation}
  \begin{split}
	K_{ab,ij} = &\frac{1}{4}\sum_{\v{k}h}d^{ab}\Big\{\chi_{ij}^{hh} + (\hat{\gamma}^z)^2\Big(\chi_{ij}^{hh} - \chi_{ij}^{h,-h}\Big)\\
	+ &\Big(\chi^{hh} - \chi^{h,-h}\Big)g_{ij}\Big\}
  \end{split}
  \label{eq:app:en:Kabij:onesum}
\end{equation}
Inserting $\chi_{ij}^{hh'}$ from Eq.~\eqref{eq:free:RhhijExpression} into this expression and preforming the approximation outlined in Eq.~\eqref{eq:free:momSumToFermAvg} for converting to Fermi surface
averages yields the expression
\begin{equation}
  \begin{split}
  K_{ab,ij} &= \frac{1}{2}\sum_h\Big\langle d^{ab}\Big\{\big[I_2^hv_i^hv_j^h - I_1^hm_{h\;ij}^{-1}\big]\\
  + &(\hat{\gamma}^z)^2\big[\big(I_2^h + I_4^h\big)v_i^hv_j^h - I_3^hv_i^{-h}v_j^h\\
  - &\big(I_1^h - I_5^h\big)m_{h\;ij}^{-1}\big] - \frac{1}{2}\big[I^h - I_0^h\big]g_{ij}\Big\}\Big\rangle_0.
  \label{eq:app:en:Kabij:integrals}
  \end{split}
\end{equation}
Here the $I$-s represent energy integrals across the energy shell around the Fermi energy of varying combinations of Green's functions as well as the density of states $N_0(\epsilon)$. As an example,
consider the integral
\begin{equation}
  \begin{split}
	I_5^h = \int_{-\epsilon_c}^{\epsilon_c}\!\!\!\mathrm{d}\xi\,\frac{N_0(\xi)}{\beta}&\sum_n\frac{1}{i\omega_n-\xi-h\abs{\v{\gamma}}}\\
	\times&\pd{}{\xi}\frac{1}{-i\omega_n-\xi+h\abs{\v{\gamma}}}.
  \end{split}
  \label{eq:app:en:I5:def}
\end{equation}
First the approximation $N_0(\xi) \approx N_F + N_F'\xi$ is used to split the integral in two: $I_5^h = I_{5,1}^h + I_{5,2}^h$, such that $I_{5,1}^h$ is the part that is proportional to $N_F$, while
$I_{5,2}^h$ is proportional to $N_F'$. The integrand of $I_{5,1}^h$ is then split using partial fractions such that
\begin{widetext}
\begin{equation}
  \begin{split}
	I_{5,1}^h &= \frac{N_F}{\beta}\sum_n\frac{1}{2(i\omega_n-h\abs{\v{\gamma}})}\int_{-\epsilon_c}^{\epsilon_c}\!\!\!\mathrm{d}\xi\bigg[\frac{1}{(-i\omega_n-\xi+h\abs{\v{\gamma}})^2} - \frac{1}{\xi^2+(\omega_n+ih\abs{\v{\gamma}})^2}\bigg]\\
	&= \frac{N_F}{\beta}\sum_n\frac{1}{2(i\omega_n-h\abs{\v{\gamma}})}\bigg[-\frac{2\epsilon_c}{\epsilon_c^2+(\omega_n+ih\abs{\v{\gamma}})^2} - \frac{2}{\omega_n+ih\abs{\v{\gamma}}}\tan^{-1}\Big(\frac{\epsilon_c}{\omega_n+ih\abs{\v{\gamma}}}\Big)\bigg]\\
	&= \frac{N_F}{\pi}\sum_n\frac{1}{i(\omega_n+ih\abs{\v{\gamma}})}\bigg[-\frac{e_c}{e_c^2+(2n+1+ih\rho)^2} - \frac{1}{2n+1+ih\rho}\tan^{-1}\Big(\frac{e_c}{2n+1+ih\rho}\Big)\bigg]\\
	&\approx -\frac{N_F\beta}{i\pi^2}\sum_n\frac{1}{(2n+1+ih\rho)^2}\tan^{-1}\Big(\frac{e_c}{2n+1+ih\rho}\Big)\\
	&\approx -\frac{\beta N_Fh}{\pi}\Im\sum_{n=0}^\infty\frac{1}{(2n+1+i\rho)^2} = -\frac{\beta N_Fh}{\pi}f_2(\rho).
  \end{split}
  \label{eq:app:en:I51}
\end{equation}
\end{widetext}
On the third line the dimensionless variables $e_c = \beta\epsilon_c/\pi$ and $\rho = \beta\abs{\v{\gamma}}/\pi$ were introduced. It was assumed that the critical temperature was low compared to the
Debye frequency such that $e_c\gg1$ and the first term on the third line could be ignored since it goes as $\sim 1/e_c$ while the arctan goes like $\sim\pi/2$. On the last line, the sum over $n$
was separated into the sum over positive and negative $n$, resulting in the imaginary component of the first sum by shifting the summation index. For $n\in[0,n_c]$, $e_c/(2n+1)\gg1$ such that
$\tan^{-1}$ is approximately $\pi/2$. $n_c$ depends on $e_c$ and since $e_c\gg1$ then $n_c\gg1$ as well such that adding the terms in the sum for $n>n_c$ doesn't change the limiting behaviour.

Similarly, the integrand of $I_{5,2}^h$ is split using partial fractions, albeit in a slightly different way which produces
\begin{widetext}
\begin{equation}
  \begin{split}
	I_{5,2}^h &= -\frac{N_F'}{2\beta}\sum_n\int_{-\epsilon_c}^{\epsilon_c}\!\!\!\mathrm{d}\xi\bigg[\frac{1}{(-i\omega_n-\xi+h\abs{\v{\gamma}})^2} + \frac{1}{\xi^2 + (\omega_n+ih\abs{\v{\gamma}})^2}\bigg]\\
	&= -\frac{N_F'}{\pi}\sum_n\bigg[\frac{1}{2n+1+ih\rho}\tan^{-1}\Big(\frac{e_c}{2n+1+ih\rho}\Big) - \frac{e_c}{(2n+1+ih\rho)^2 + e_c^2}\bigg]\\
	&\approx -\frac{2N_F'}{\pi}\Re\sum_{n=0}^\infty\bigg[\frac{\tan^{-1}\Big(\frac{e_c}{2n+1+i\rho}\Big)}{2n+1+i\rho} - \frac{\tan^{-1}\Big(\frac{e_c}{2n+1}\Big)}{2n+1} + \frac{\tan^{-1}\Big(\frac{e_c}{2n+1}\Big)}{2n+1}\bigg]\\
	&\approx -N_F'f(\rho) -\frac{N_F'}{2}\ln(2e_ce^C).
  \end{split}
  \label{eq:app:en:I52}
\end{equation}
\end{widetext}
Inserting these results back into Eq.~\eqref{eq:app:en:I5:def} then yields
\begin{equation}
  I_5^h \approx -\frac{N_F\beta h}{\pi}f_2(\rho) -N_F'f(\rho) - \frac{N_F'}{2}\ln(2e_ce^C).
  \label{eq:app:en:I5:calc}
\end{equation}
The remaining integrals are calculated in a similar manner. In the cases where $\rho/e_c$ remains in the expression after integrating, this is expanded to first order in $O(\rho/e_c)$, e.g. in $I_0^h$. Terms
proportional to $e^{-e_c}$ are also neglected like in $I_2^h$. With these approximations the integrals become
\begin{widetext}
\begin{subequations}
  \begin{align}
	I^h &= \int_{-\epsilon_c}^{\epsilon_c}\!\!\!\!\mathrm{d}\xi\,\frac{N_0(\xi)}{\beta}\sum_n\frac{1}{\omega_n^2 + (\xi+h\abs{\v{\gamma}})^2} 
	\approx N_F\ln\big(2e_ce^C\big) + hN_F'\abs{\v{\gamma}}\big[1-\ln\big(2e_ce^C\big)\big], \label{eq:app:en:int:I}\\
	I_0^h &= \int_{-\epsilon_c}^{\epsilon_c}\!\!\!\!\mathrm{d}\xi\,\frac{N_0(\xi)}{\beta}\sum_n\frac{1}{\xi^2 + (\omega_n+ih\abs{\v{\gamma}})^2}
	\approx N_F\ln\big(2e_ce^C\big) + 2N_Ff(\rho), \label{eq:app:en:int:I0}\\
	I_1^h &= \int_{-\epsilon_c}^{\epsilon_c}\!\!\!\!\mathrm{d}\xi\,\frac{N_0(\xi)}{\beta}\sum_n\frac{1}{i\omega_n-\xi-h\abs{\v{\gamma}}}\pd{}{\xi}\frac{1}{-i\omega_n-\xi-h\abs{\v{\gamma}}} 
	\approx -\frac{N_Fh\abs{\v{\gamma}}}{2\epsilon_c^2} - \frac{N_F'\ln\big(2e_ce^C\big)}{2}, \label{eq:app:en:int:I1}\\
	\begin{split}
	  I_2^h &= \int_{-\epsilon_c}^{\epsilon_c}\!\!\!\!\mathrm{d}\xi\,\frac{N_0(\xi)}{\beta}\sum_n\bigg({\textstyle\pd{}{\xi}\frac{1}{i\omega_n-\xi-h\abs{\v{\gamma}}}\pd{ }{\xi}\frac{1}{-i\omega_n-\xi-h\abs{\v{\gamma}}} - \frac{1}{i\omega_n-\xi-h\abs{\v{\gamma}}}\pdd{ }{\xi}\frac{1}{-i\omega_n-\xi-h\abs{\v{\gamma}}}}\bigg)\\
	  &\approx\frac{7\zeta(3)\beta^2}{4\pi^2}(N_F-h\abs{\v{\gamma}}N_F'), \label{eq:app:en:int:I2}
	\end{split}\\
	I_3^h &= \int_{-\epsilon_c}^{\epsilon_c}\!\!\!\!\mathrm{d}\xi\,\frac{N_0(\xi)}{\beta}\sum_n\pd{}{\xi}\frac{1}{i\omega_n-\xi-h\abs{\v{\gamma}}}\pd{}{\xi}\frac{1}{-i\omega_n-\xi+h\abs{\v{\gamma}}}
	\approx \frac{N_F\beta^2}{\pi^2}\Big(f_3(\rho) + \frac{7\zeta(3)}{8}\Big), \label{eq:app:en:int:I3}\\
	\begin{split}
	  I_4^h &= \int_{-\epsilon_c}^{\epsilon_c}\!\!\!\!\mathrm{d}\xi\,\frac{N_0(\xi)}{\beta}\sum_n\frac{1}{i\omega_n-\xi-h\abs{\v{\gamma}}}\pdd{}{\xi}\frac{1}{-i\omega_n-\xi+h\abs{\v{\gamma}}}\\
	  &\approx -\frac{N_F\beta^2}{\pi^2}\Big(f_3(\rho) + \frac{7\zeta(3)}{8}\Big) + \frac{N_F'\beta h}{\pi}f_2(\rho), \label{eq:app:en:int:I4}
	\end{split}\\
	\begin{split}
	  I_5^h &= \int_{-\epsilon_c}^{\epsilon_c}\!\!\!\!\mathrm{d}\xi\,\frac{N_0(\xi)}{\beta}\sum_n\frac{1}{i\omega_n-\xi-h\abs{\v{\gamma}}}\pd{}{\xi}\frac{1}{-i\omega_n-\xi+h\abs{\v{\gamma}}}\\
	  &\approx -\frac{N_F\beta h}{\pi}f_2(\rho) - \frac{N_F'\ln\big(2e_ce^C\big)}{2} - N_F'f(\rho). \label{eq:app:en:int:I5}
	\end{split}
  \end{align}
  \label{eq:app:en:int}
\end{subequations}
\end{widetext}
The expression for $K_{ab,ij}$ in Eq.~\eqref{eq:free:Kab:FermSurfAverages} is then obtained by inserting these integrals into Eq.~\eqref{eq:app:en:Kabij:integrals} and summing over $h$.
The integrals $I^h$ and $I_0^h$ are used to obtain the expression for $A_{ab}$ in Eq.~\eqref{eq:free:Aab:FermiAvg}.
\bibliography{probib}

\end{document}